\documentclass[elsarticle,floatfix,superscriptaddress]{revtex4}
\usepackage{graphicx}
\usepackage{braket}
\usepackage[caption=false]{subfig}
\usepackage{amssymb}
\usepackage{epstopdf}
\usepackage{natbib}

\newcommand{\bea}{\begin{eqnarray}}
	\newcommand{\eea}{\end{eqnarray}}
\newcommand{\beq}{\begin{equation}}
	\newcommand{\eeq}{\end{equation}}
\usepackage{color}

\begin{document}

\title{A topological Josephson junction platform for creating, manipulating, and braiding Majorana bound states}

\author{Suraj S. Hegde}
\affiliation{Department of Physics, University of Illinois at
Urbana-Champaign, Urbana, Illinois, USA}
\affiliation{Max Planck Institute for Physics of Complex Systems, Dresden, Germany.}
\author{Guang Yue}
\affiliation{Department of Physics, University of Illinois at
Urbana-Champaign, Urbana, Illinois, USA}
\author{Yuxuan Wang}
\affiliation{Department of Physics,University of Florida, Gainesville, Florida, USA}
\affiliation{Department of Physics, Stanford University, California, USA}
\author{Erik Huemiller}
\author{D. J. Van Harlingen}
\author{Smitha Vishveshwara}
\affiliation{Department of Physics, University of Illinois at
Urbana-Champaign, Urbana, Illinois, USA}
\email[shegde2@illinois.edu]{Your e-mail address}


\begin{abstract}
As part of the intense effort towards identifying platforms in which Majorana bound states can be realized and manipulated to perform qubit operations, we propose a topological Josephson junction architecture that achieves these capabilities and which can be experimentally implemented.  The platform uses conventional superconducting electrodes deposited on a topological insulator film to form networks of proximity-coupled lateral Josephson junctions. Magnetic fields threading the network of junction barriers create Josephson vortices that host Majorana bound states localized in the junction where the local phase difference is an odd multiple of $\pi$, i.e. attached to the cores of the Josephson vortices.  This enables us to manipulate the Majorana states by moving the Josephson vortices, achieving functionality exclusive to these systems in contrast to others, such as those composed of topological superconductor nanowires.  We describe protocols for: 1) braiding localized Majorana states by exchange, 2) controlling the separation and hence the coupling of adjacent localized Majorana states to effect non-Abelian rotations via hybridization of the Majorana modes, and 3) reading out changes in the non-local parity correlations induced by such operations. These schemes utilizes current pulses and local magnetic field pulses to control the location of vortices, and measurements of the Josephson current-phase relation to reveal the presence of the Majorana bound states. Finally,we present brief discussions of readout schemes and viable experimental settings for realizing the platform. 
\end{abstract}


\maketitle

\section{Introduction}
\label{sec:Introduction}

Topologically-protected quantum processing is rapidly emerging as a viable route to next generation advances in quantum information, computational science and technology \citep{Nayak08, Alicea11,Pachos,Field}.  In contrast to conventional qubits, topological qubits are based on exotic quasiparticle excitations in condensed matter systems that exhibit non-Abelian statistics. Systems having anyons obeying braiding rules are expected to show resilience to environmental interference, making them excellent candidates for fault-tolerant quantum computation \citep{KitaevQC}. Within the last decade, the proposal and subsequent evidence for the experimental realization of topological superconductivity capable of hosting Majorana bound states (MBS) has created intense attention and activity from the perspective of such topological qubit technology \cite{Alicea11,Choy11,Kjaergaard12,Pientka13,Martin12,Das12,Churchill13,Rokhinson12,Deng12,Finck13}. In these fermionic systems, these localized Andreev bound states are predicted to exist as zero energy states at the Fermi energy.  A non-local pair of such MBSs share an electronic state that can either be occupied or empty, making such a pair a parity qubit. Implementing qubit operations requires positioning and manipulating MBSs and exploiting their non-Abelian nature through braiding\cite{Ivanov01,Alicea12}. While such MBSs cannot alone span universal quantum computation, they are currently the forerunners for achieving topological qubits\cite{Hassler14,Sau10}. Critical steps required for realizing topological quantum processing are under development, including experimental verification of the existence of MBSs, creating architectures that offer a platform for qubit operations, and designing complex non-Abelian braiding-based quantum computational protocols.

As with conventional qubits, now realized in a wide range of systems including coupled spins, superconducting transmons, photonic circuits, and cold atom systems \cite{Ginossar14,Jiang11,Xu16}, it is imperative that multiple promising approaches be explored to optimize progress toward successful implementation of topological quantum computing schemes. The past years have indeed witnessed a growth of potential candidate systems for hosting topological qubits mainly centered around MBSs \cite{Das12,Churchill13,Rokhinson12,Deng12,Finck13,Mourik12}. Most attention has been on nanowires having strong spin-orbit coupling and proximitized by contact with a conventional superconductor.  By applying a magnetic field oriented along the length of the wire, this system can be tuned into a topological state in which MBSs are predicted to nucleate at the ends of the wire. More recently, chains of ferromagnetic atoms fabricated on a superconducting surface \citep{Nadj-Perge14}, have received prominent attention for their ability to host MBSs, also attached to the ends of the wire. Novel materials that exhibit quantum Hall physics, such as graphene, have revived the initial interest of over a decade ago of exploiting certain fractional states, such as $\nu=5/2$, for their potential to harbor Majorana bound states, as well as other states having more exotic fractional quasiparticle excitations (e.g. parafermions) that can, in principle, perform universal quantum computation\cite{Hutter16}. 

\begin{figure}
\includegraphics[width=0.5\columnwidth]{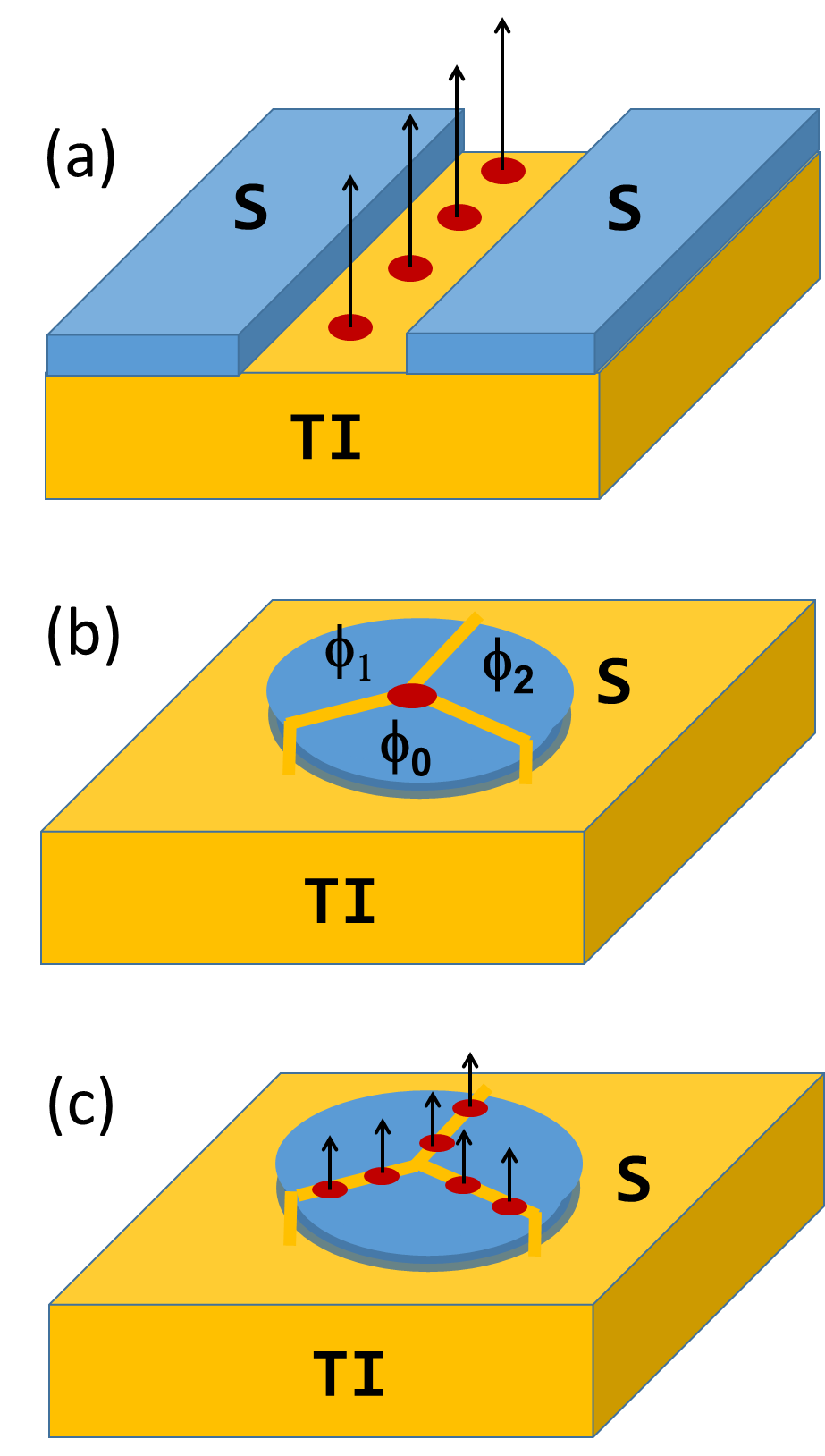}
\caption{Nucleation of Majorana bound states in S-TI-S structures:  (a) Lateral S-TI-S Josephson junction in a magnetic field with MBSs at the locations of Josephson vortices, (b) trijunction in zero magnetic field with a single MBS in the center induced by appropriate adjustment of the phases on the electrodes, and (c) trijunction in a magnetic field with multiple MBSs.} \smallskip
\label{fig:1}
\vspace{-1pt}
\end{figure}

As a viable alternative to these systems in which the topological excitations of interest are physically bound to the end or edges of 1D or 2D structures, here we propose a platform (Fig.\ref{fig:1}) consisting of multiply-connected lateral Superconductor-Topological Insulator-Superconductor (S-TI-S) Josephson junctions networks for realizing MBSs whose locations can be controllably moved, providing additional functionality for braiding and hybridization that perform non-Abelian operations \cite{Fu08,Fu09,Potter13,Grosfeld11,Stern19,Zwan15,Pientka19,Schrade15,Volpez19,Hell17,Xie19,sakurai20,vanHeck12,Zhou20,Setiawan19,Stern19,Graziano20,Serra20}. In this paper, we delineate the key features required to realize these MBSs and carry out topological quantum processes in this system. Our proposed lateral S-TI-S Josephson junctions offer an attractive platform for MBS manipulations for the following reasons:  (1) MBSs in this system are zero-energy Andreev bound states enabled by the spin-momentum locking of topological surface states in the TI and stabilized by the phase of the Josephson coupling.  (2) In contrast to other systems such as semiconductor nanowires\cite{Mourik12}, nucleation of the MBSs does not require a large magnetic field, enabling phase-sensitive Josephson measurements. (3) Magnetic fields instead play a different role by localizing MBSs at Josephson vortex cores, which allows us to move the MBSs by moving the vortices, easily done in controlled ways by applying currents, voltages, or phase differences. (4) The MBSs can be created in a controlled way in uniform junction regions and are not subject interface issues, unlike with nanowires in which the MBSs exist at interface between topological and non-topological regions.   (5) Junction networks are easily scalable to create complex circuits, and surface codes for performing universal quantum computing in networks of Josephson junctions have already been proposed \citep{vanHeck12,deLange15,Vijay15}. As we briefly review in a later section, there have already been extensive measurements of the transport and Josephson properties of S-TI-S junctions and many of the features expected to result from MBSs in this system have been observed \cite{Stehno16, Kurter14, Kurter15, Sochnikov13,Sochnikov15,ExperimentPaper}.

In order to show the suitability of these S-TI-S platforms for the purposes of quantum processing, it is required to explicitly demonstrate certain benchmark tasks. First and foremost, one requires the nucleation of MBSs and their detection of through various local probes (tunnel junctions, quantum dots, single-electron transistors) and interferometry techniques (critical current diffraction patterns, current-phase relation measurements). Here, threading magnetic flux through Josephson junctions and extracting the critical current modulation patterns provides a natural means for realizing both these aspects. Second, one requires the measurement of parity states encoded in pairs of these Majorana modes. The platform proposed here, in addition to allowing for charge sensitive measurements, such as proposed in the nanowire case, can reveal parity transitions through switches in the relative sign of the 4$\pi$-periodic component of the Josephson current-phase relation that arises from MBS currents. Finally, non-Abelian rotations in the ground state manifold require the manipulation of at least four Majorana modes. Though this can be strictly shown through actual braiding of Majorana modes through their motion in physical space \citep{Ivanov01},  a simpler set-up performs this in a non-universal way involves bringing two of the MBSs within each other's proximity \citep{Burnell14, Bonderson08, Bonderson13,Chiu15,Sau11}. Within the proposed  architecture of networks formed by Josephson junctions, the motion of MBSs bound to  Josephson vortices enables both kinds of braiding (exchange and hybridization). 
 
 The purpose of this work is to present an appropriate topological Josephson junction architecture that can demonstrate all these tasks crucial for a functional MBS-based quantum processing platform. By bringing together theory and experimental expertise, we design and model principles for nucleating and braiding MBSs in realistic geometries formed of the best candidate materials, informed by experiments being performed in tandem by some of the authors\cite{ExperimentPaper} towards realizing the first steps of the design. In what follows, in Section II, we begin with a short summary of what the platform entails. In Section III, we provide a theoretical modeling of the extended topological Josephson junction, focusing on an effective one-dimensional description of the two dispersive Majorana states at the S-TI-S interfaces \cite{Potter13}. We demonstrate and analyze cases where the application of flux results in multiple zero energy MBSs formed by localizing the dispersive modes along the junction.  In Section IV, we calculate the modulation of the critical current as a function of applied flux that is sensitive to the interference of the Josephson supercurrents in the junction, which reveals the effect of the Majorana modes, specifically, node-lifting of odd nodes in comparison with the Fraunhofer diffraction patterns expected for uniform junctions with the usual sinusoidal current-phase relation.  In Section V, we present schemes for performing non-Abelian rotations of the MBSs via exchange and hybridization. In Section VI, we discuss implementating these exchange and hybridization operations in realistic experimental systems in terms of circuit and applications of current and field pulses. In Section VII, borrowing from lessons learned in the case of nanowires, we outline possible schemes for readout of Majorana qubit states. We finally comment on the experimental status and viability of this platform in Section VIII and conclude in Section VIII with a recapitulation of our proposal and findings in the context of the broader outlook for topological quantum processing. 

\section{S-TI-S Josephson junctions as a platform for Majorana bound state nucleation and manipulation}

Here we provide an overview of the proposed S-TI-S Josephson junction platform for nucleating and manipulating Majorana bound states (MBSs).\smallskip

{\bf Platform architecture:}  The basic S-TI-S Josephson junction building blocks are made of superconducting islands deposited on top of a topological insulator to form single junctions and trijunctions consisting of three superconducting regions adjacent to each other, as shown in Figure \ref{fig:1}. More complex architectures consisting of appropriate networks of junctions can be constructed from these building blocks. For instance, a typical repetitive network pattern could take on a honeycomb structure consisting of a lattice of hexagonal shaped superconducting regions. The architecture would integrate leads, electrodes, single-electron transistors, or microwave cavities depending on the manipulation and read-out schemes. \smallskip

\begin{figure}[htp]
\includegraphics[width=0.7
\columnwidth]{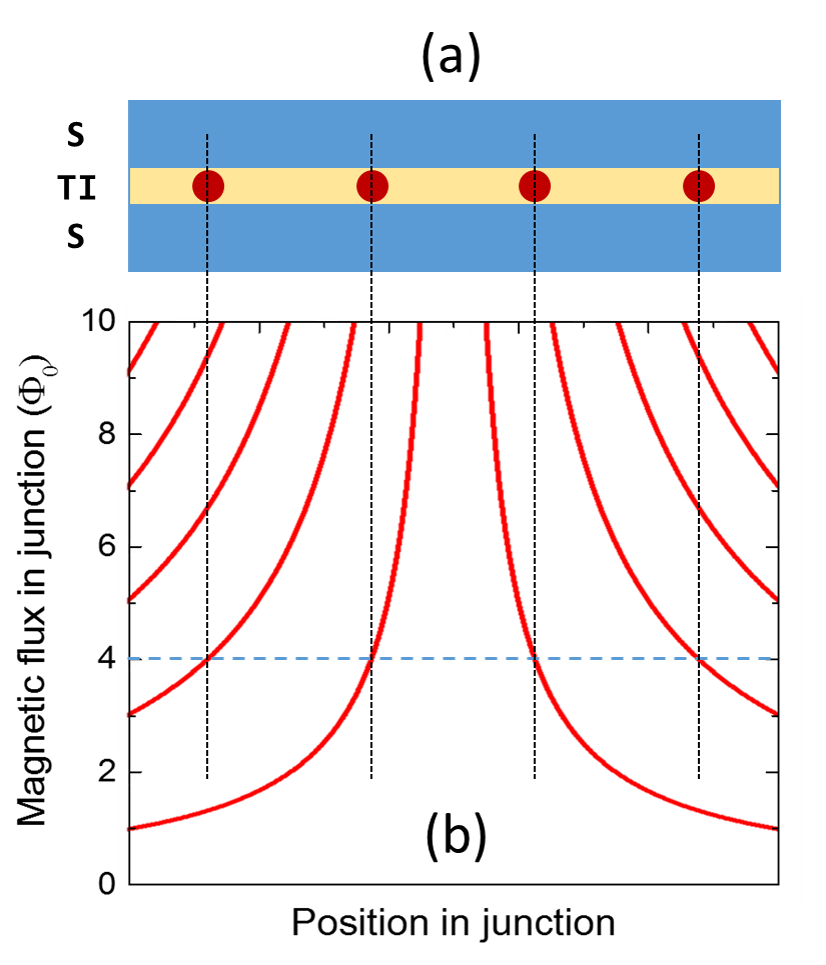}
\caption{(a) Location of Josephson vortices and the Majorana states bound to them at an applied magnetic flux of  $\Phi_0$ in the junction. (b) Location of vortices as a function of the applied magnetic flux. At zero current, the MBSs enter the junction symmetrically from each side as the flux is increased, first entering when there is one flux quantum in the junction.}
\label{fig:2}
\end{figure}

 {\bf Nucleating and identifying Majorana bound states:}  A simple method for nucleating MBSs involves applying a magnetic flux through a Josephson junction.  Along the extended junction line, the phase difference across the junction depends both on the applied field and on forced condensate phases which can be controlled externally. The applied field induces a gradient in the phase across the junction and a non-uniform supercurrent in the junction.  As the magnetic field is increased, Josephson vortices enter the junction, symmetrically from each side of the junction at zero applied current, with cores located at the points where the relative phase between the two superconductors is equal to $\pi$ or an odd multiple thereof.  These vortices are evenly-spaced in a uniform junction, separated by a flux of one $\Phi_0$ threading the junction barrier, as shown in Figure \ref{fig:2}. Localized MBS are stabilized at these points, effectively bound to the Josephson vortices \cite{Fu08,Potter13}. Such bound states will form the basis of our proposed schemes. While applying a relative phase shift of $\pi$ between the three superconducting regions in the trijunction geometry of Figure \ref{fig:1}(b) can also nucleate a MBS at the intersection of its junctions, such states will not be our focus here \cite{Fu08}. As signatures of the flux-induced appearance of MBSs in extended junction geometries, we will show that the critical current diffraction patterns, which track critical current as a function of applied flux, exhibit characteristic features due to low-energy Majorana-mode contributions to the Josephson critical current.  
 
\begin{figure}
\includegraphics[width=1.0
\columnwidth]{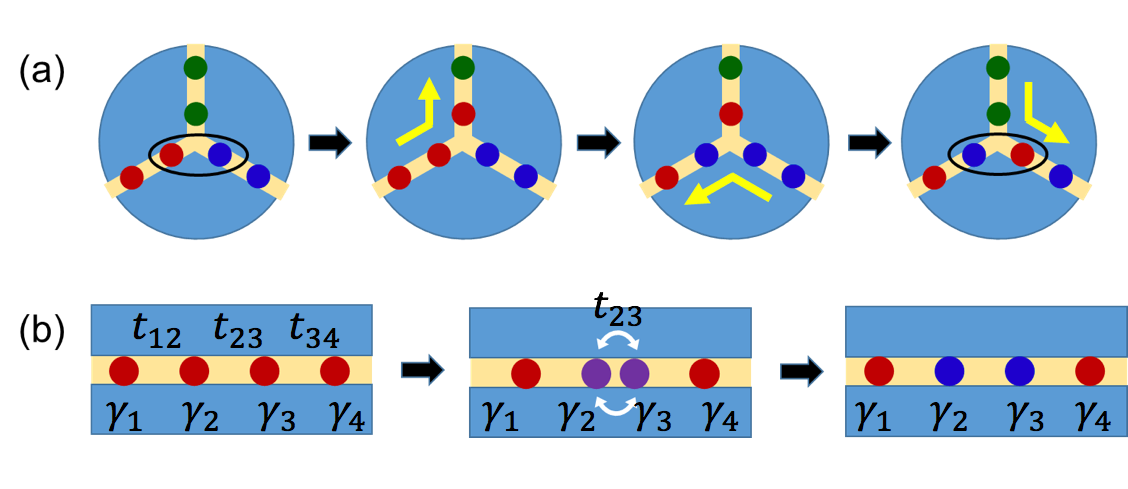}
\caption{Braiding schemes: (a) Braiding by exchange of MBSs in a trijunction via successive movement of vortices/MBSs along different arms of the trijunction, resulting in an exchange of two MBSs, and  (b) effective braiding by hybridization moving adjacent MBSs close together to induce coupling that evolves their relative phase. The MBSs are denoted by $\gamma_1$,$\gamma_2$, $\gamma_3$ and $\gamma_4$. The coupling between the pairs of  MBS $(\gamma_i,\gamma_j)$ are denoted by $t_{ij}$.}
\label{fig:3}
\end{figure}

  {\bf Non-Abelian rotations via braiding and hybridization:} Once the Majorana modes are nucleated the next task is to perform  non-Abelian rotations in the ground state manifold. Pairs of MBSs define electronic parity states which can either be occupied or not. The rotations are in this Hilbert space. We propose two schemes for performing rotation based on a theoretical framework developed to describe the localization of Majorana bound states in a magnetic field and their manipulation by local fields and currents. The first approach relies on applying a series of phase pulses in a trijunction S-TI-S device resulting in the exchange of Josephson vortices containing MBSs, resulting in braiding (Figure \ref{fig:3}(a)).  The second uses magnetic field pulses to control the separation of vortices in a single Josephson junction (Figure \ref{fig:3}(b)), resulting in hybridization of MBSs that creates an energy splitting away from zero energy and an associated rotation in the Hilbert space of parity states.  This scheme lacks the full topological protection of braiding, but is highly implementable in our geometry.  \smallskip

{\bf Parity readout:}  Either as a means of initializing parity states or doing readouts, determining the parity of electronic states shared by MBS pairs is a crucial ingredient. In the situation here, Josephson junction physics provides a natural way of determining such non-local parity--critical current switching measurements  \citep{Peng16}. As in the nanowire situation, other options involve coupling to quantum dot single-electron transistors (SET) or embedding the platform in a transmon geometry.  In this work, we briefly survey the possibilities based on the capabilities of the Josephson junction architecture. \smallskip 

In the next section, we describe the theoretical principles behind modeling these S-TI-S junctions, obtaining critical current modulation patterns, and performing non-Abelian rotations.

\section{Modeling S-TI-S junctions}
\label{sec:model}
Here we begin our extensive treatment of the proposed S-TI-S architecture by describing a single extended Josephson junction and the low-energy dispersive Majorana modes that reside on the TI surface in the proximitized region between the superconductor electrodes. We review the manner in which localized MBS nucleate in the presence of applied flux. We then analyze in depth situations having multiple MBSs, calculating their energy spectra, wavefunctions, effect of a non-uniform magnetic flux, and hybridization arising from tunnel coupling of the Majorana modes.

\subsection{Effective model of low-energy junction modes}
\begin{figure}
\includegraphics[width=1.0
\columnwidth]{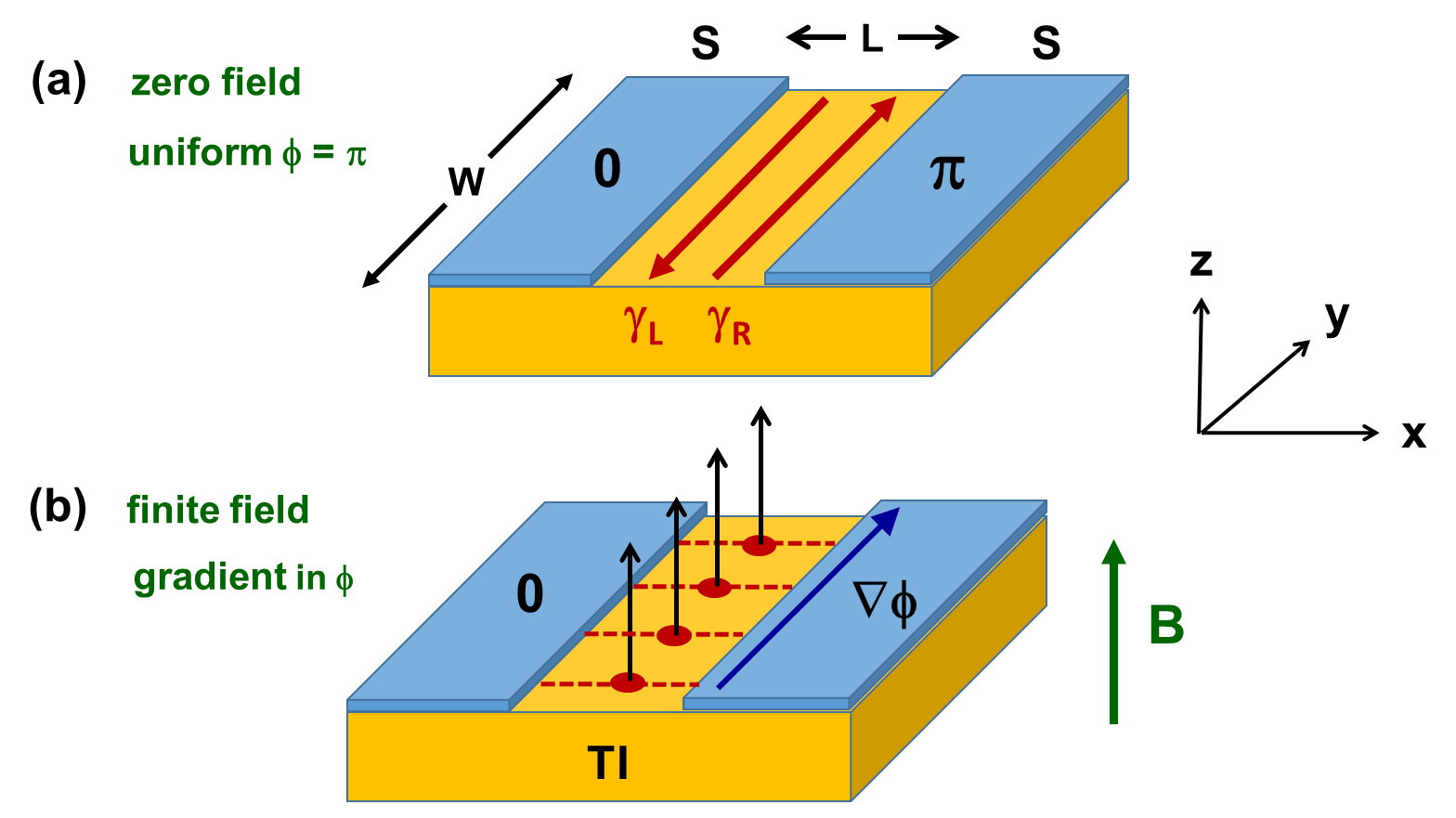}
\caption{Basic features of the Josephson junction formed by two superconducting regions deposited onto the top surface of a topological insulator.  The width of the junction, oriented along the $y$-axis, is of dimension $W$, and its length between the electrodes, along the $x$-axis, is of dimension $L$.  (a)  In zero magnetic field and a uniform phase difference of $\pi$ across the junction, this structure supports low-energy counter-propagating dispersing Majorana modes $\gamma_L$ and $\gamma_R$ in the barrier on the surface.   (b) Applied magnetic flux through the junction creates a spatial gradient in the phase difference between the superconductors and, if sufficiently large, generates Josephson vortices that localize the Majorana modes into discrete Majorana bound states pairs at locations where the phase difference is an odd-multiple of $\pi$.  The red dots indicate the location of the localized MBSs on the top surface of the junction. The other MBS of the pair is delocalized in this geometry.  The situation is modeled by Eq. \ref{Eq:Ham}.}
\label{fig:4}
\end{figure}

The basic structure of the junctions studied here is as shown in Figure \ref{fig:4}. It consists of a topological insulator slab with two superconducting islands deposited on its upper surface. We assume here that the TI slab is much thicker than coherence length of the proximity-induced superconductivity so that the Josephson supercurrents are confined to the top surface. We restrict ourselves to a well-established effective model that focuses on the low-energy states found along the junction interfaces \cite{Potter13}.

Let us consider a S-TI-S system, similar to the one described in Ref. \onlinecite{Potter13} having a line junction of width $W$ along $y$-axis as shown in Figure \ref{fig:4}. The superconducting gap varies as : $\Delta(x)= \Delta e^{i\phi(y)}$ for $x>L/2$ and $\Delta(x)=\Delta $ for $x<-L/2$. A magnetic field pierces through this junction with flux $\Phi$. The flux leads to a spatial variation of the superconducting phase difference along the junction to vary as 
\beq
\phi(y)=2\pi y/l_B, 
\eeq
where $l_B=W \Phi_0/\Phi$. Here $\Phi_0=h/2e$ is the flux quantum appropriate for paired superconductivity.

It can be shown \cite{Potter13} when the two SC islands are decoupled, at each S-TI interface there exists a dispersive Majorana mode at zero field,  thus yielding a pair of counter-propagating states $\gamma_{L}$ and $\gamma_{R}$ along the junction, as illustrated in Figure \ref{fig:4}(a). The desired MBSs are particular Andreev bound states that couple these modes and form localized states in the presence of a vertical field of sufficient magnitude, as shown in Figure \ref{fig:4}(b).  

The low-energy effective Hamiltonian describing the situation has the form
\beq
H_{eff}=i v_M (\gamma_L \partial_y \gamma_L - \gamma_R \partial_y \gamma_R)+i \Delta \cos(\phi(y)/2)\gamma_L\gamma_R
\label{Eq:Ham}
\eeq
with $v_M = v [\cos(\frac{\mu W}{  v})+(\frac{\Delta}{ \mu})\sin(\frac{\mu W}{  v})]\frac{\Delta^2}{(\mu^2 + \Delta^2)}$, where $v$ is the velocity corresponding to the edge state of the TI  and $\mu$ is the chemical potential. For a S-TI-S junction of Al-Bi$_2$Se$_3$-Al the estimated values are $v=10^5ms^{-1}$, $\Delta=150 \mu eV$, $\mu=10 meV$ \citep{Fu08, Potter13}. The energy gap is an order of magnitude larger for Nb electrodes often used in experiments.

The form of Eq. \ref{Eq:Ham} respects the Dirac equation for a massive particle, where the gap function $ \Delta \cos(\phi(y)/2)$ represents a spatially varying mass function. For a linear variation of the flux-dependent $\phi(y)$, the gap function too can be linearized around regions where $\phi(y)$ crosses an odd integer multiple of $\pi$. In this case, there exists a zero-energy eigenstate that shows exponential decay away from the crossing point \cite{Fu08,Grosfeld11}. This eigenstate has the appropriate linear combination of $\gamma_R$ and $\gamma_L$ such that the desired Majorana bound states (MBSs) are real functions.  As the magnetic field piercing through the junction is increased, the number of zeros of the gap function increases, thus capturing more number of Majorana modes in the junction. A new Majorana mode appears with the incremental change of the net flux by one quantum, thus confining one Majorana bound state per one Josephson vortex.  

A few comments are in order here with regards to several simplifying assumptions made in this model. Here we assume the dimension $L$ to be small. That ensures that the tunneling is local and directional within an extended junction, a requirement for using Josephson interferometry probe the phase variations in the junction.    We also assume that the junctions are in the Josephson "short-junction limit" in which the magnetic fields produced by the supercurrents are small compared to the applied field. This a good assumption for the S-TI-S we study as a result of the small Josephson current density.  In extended junctions with larger current density in the "long-junction limit", the phase dynamics of Josephson fluxons must be considered and the the gap function can exhibit Josephson soliton states with a $tanh$-like spatial variation. Here too, respecting the generic change of sign in the gap function, it can be shown that there exists a Majorana bound state\citep{Grosfeld11}. 

A significant issue is that the physical Hilbert space requires that the MBSs appear in pairs while in our continuum model on  the  surface,  it  is  possible  to  obtain  a  single  MBS. In Ref. \onlinecite{Potter13}, the junctions treated are symmetric with superconducting electrodes on both the top and bottom surface of the TI.  In that case, pairs of MBS exist on each surface of the TI. For this to hold, the induced superconducting penetration length within the TI must be greater than its thickness. Here, we consider the situation in which the superconductor electrodes are only on the top surface. For this system, there are localized MBS on on that surface and its partner is delocalized, likely extended along the periphery of the superconducting islands.

The effective low-energy theory thus gives us an excellent starting point for modeling the primary phenomena and protocols described in this work. Future work would build on each of these aspects and factors described above in a more detailed way by employing a full-fledged three-dimensional model. As the simplest next step, the model could involve a lattice version of a Bogoliubov-de Gennes Hamltonian describing the TI and  proximity-induced SC hybrid geometry (where as a next step, the induced superconductivity could be self-consistently evaluated). Such a model would show the exact distribution of the low-lying dispersing modes of relevance in space and energy. Such a model would reveal a more precise map of the MBS locations, other low-energy bound states, and values of the tunnel coupling that are more potentially realistic compared to the estimates we present in what follows. We now turn to a detailed analysis of the MBS within the context of the effective low-energy model.

\subsection{Derivation of the effective model}
Here, so as to make our presentation self-contained, we recapitulate the derivation of the effective model  Eq.\ref{Eq:Ham} as originally presented in  Ref.\cite{Fu08}. Consider the model for the surface of a topological insulator with proximity induced superconductivity. The second quantized Hamiltonian is given by $H=\Psi^{\dagger}\mathcal{H}\Psi/2$
\begin{equation}
    \mathcal{H} = -iv\tau_z \sigma_y \partial_y -i v \tau_z \sigma_x \partial_x -\tau_z \sigma_0 \mu +\tau_x \sigma_0 \Delta \cos \phi+\tau_y \sigma_0 \Delta \sin \phi
    \label{Eq:TISC Ham}
\end{equation}
Superconductivity is treated at the mean field approximation and is characterized by the pairing amplitude $\Delta$. The Nambu basis $\Psi$ is given by $\Psi =((\psi_{\uparrow}, \psi_{\downarrow}),i\sigma_y(\psi^{\dagger}_{\uparrow},\psi^{\dagger}_{\downarrow}))^T$. The Pauli matrices $\sigma,\tau$ corresponds to spin and particle-hole degrees of freedom. 

To obtain the effective model, first consider the configuration in Fig.\ref{fig:4}a and solve for the zero-energy chiral Majorana modes which propagate along the x-direction. The Hamiltonian above can then be projected onto the basis of those zero modes to obtain the effective model. One can choose the solutions to be simultaneous eigensolutions of particle-hole conjugation operator $\sigma_y \tau_y K$, where $K$ is the complex conjugation:$\ket{\zeta_{\pm}} = \frac{1}{2}(1,\pm i,\pm i,-1)^T = \ket{\zeta_1} \pm i \ket{\zeta_2}$. The zero-energy solutions are given by
\begin{equation}
    \Psi^0_{\pm} = \frac{1}{2}(1,\pm i,\pm i,-1)^T e^{\pm i \mu x/v}e^{-\int^{|x|}_0 \Delta(x')/v  dx'}
\end{equation}
The two independent real Majorana modes can be obtained from the above by taking the real and imaginary parts
\begin{equation}
    \bar{\gamma}_L = \exp\bigg( \frac{1}{v}\int^x \Delta(x') dx' \bigg)(\cos(\mu x/v) \ket{\zeta_1} + \sin(\mu x/v) \ket{\zeta_2}) \gamma_L(y)
\end{equation}
\begin{equation}
    \bar{\gamma}_R = \exp\bigg(\frac{1}{v}\int^x \Delta(x') dx'\bigg) (-\sin(\mu x/v) \ket{\zeta_1} + \cos(\mu x/v) \ket{\zeta_2})\gamma_R(y)
\end{equation}

The effective model can thus be obtained by projecting the Hamiltonian $\mathcal{H}$ to the basis of the chiral Majorana modes $\bar{\gamma}^T= (\bar{\gamma}_L,\bar{\gamma}_R)^T$:
\begin{equation}
    H_{eff}=\int dy \bar{\gamma}^T \tilde{\mathcal{H}} \bar{\gamma}
\end{equation}
The projection involves taking the following overlaps of the Pauli matrices that appear in the Hamiltonian Eq.\ref{Eq:TISC Ham}:
\begin{eqnarray}
\bra{\zeta_1}\tau_z \sigma_y\ket{\zeta_2}= \bra{\zeta_2}\tau_z \sigma_y\ket{\zeta_1}=1 \\
\bra{\zeta_{1,2}}\tau_z \sigma_y\ket{\zeta_{1,2}}=0 \\
-\bra{\zeta_1}\tau_y \sigma_0 \ket{\zeta_2}=\bra{\zeta_2}\tau_y \sigma_0\ket{\zeta_1}=i \\
\bra{\zeta_{1,2}}\tau_y \sigma_0\ket{\zeta_{1,2}}=0
\end{eqnarray}

The first term in Eq. \ref{Eq:TISC Ham} projects to 
\begin{equation}
    \int dx dy \bar{\gamma}^T (-iv \tau_z \sigma_y\partial_y) \bar{\gamma}= \int dy  iv_M ( \gamma_L \partial_y \gamma_L-\gamma_R\partial_y \gamma_R),
\end{equation}
where we have introduced the $\gamma^T=(\gamma_L(y),\gamma_R(y))^T$. The `renormalised' velocity is given by
\begin{equation}
    v_M= \int dx e^{-2\int_0^x \frac{\Delta(x')}{v}dx'} \cos(\frac{2\mu x}{v})= v\frac{\Delta^2}{\mu^2 + \Delta^2} \bigg( \cos \bigg(\frac{\mu L}{v} \bigg)+ \frac{\Delta}{\mu}\sin \bigg(\frac{\mu L}{v} \bigg) \bigg)
\end{equation}

The projection of the superconducting term  gives rise to a mass term $m(y)$ for the Majorana modes
\begin{equation}
    \int dx dy \bar{\gamma}^T (\tau_x \sigma_0 \Delta \cos \phi+\tau_y \sigma_0 \Delta \sin \phi) \bar{\gamma}= i\int dy (m(y) \gamma_L \gamma_R),
\end{equation}
where the mass term is given by
\begin{equation}
    m = \int dx \Delta(x) \sin (\phi) \exp\bigg( \frac{2}{v} \int_0^x dx' \Delta(x) \cos(\phi(x')) \bigg) \propto -\Delta \cos \bigg( \frac{\phi}{2} \bigg)
\end{equation}

The final effective Hamiltonian thus takes the form 
\begin{equation}
    \mathcal{H}_{eff} =-iv_M \tilde{\tau}_x \partial_y + \tilde{\tau}_y m(y)
\end{equation}

 This low-energy description has the same structure as that of the effective model for a Su-Schrieffer-Heeger model. Here the Pauli matrices $\tilde{\tau}$ are in the basis of $\gamma^T=(\gamma_L,\gamma_R)^T$. As known from the Jackiw-Rebbi problem, when mass term has a spatial profile such that it smoothly changes sign (as in a soliton profile),  there is a topologically protected zero-mode localised at that point. In our case, the mass term is $m=\cos(\phi/2)$ and therefore flips sign at $\phi=\text{odd multiples of }\pi$. One thus obtains isolated, localised Majorana zero modes at such vortices or phase slips. The spatial profile of this mode is of the form:
 \begin{equation}
     \gamma_0 \sim  e^{-\int m(y')dy'} (1,0)^T
 \end{equation}
 
  As we have explained above, application of a magnetic field perpendicular to the junction leads to a linear variation of the SC phase $\phi= 2\pi y/ \ell_B$ and creates points of phase slips/vortices as the magnetic field is increased. Majorana bound states are thus localized at these phase slip points.

 \subsection{Multiple vortices and numerical analysis}
Here we analyze the situation in which the applied flux is strong enough to generate multiple vortices and MBSs. In particular, we study the case of four MBSs present along the junction; such a situation is the minimum necessary for quantum information protocols. Through numerical simulation of the model presented in the sub-section above, we show the explicit realization of these MBS states, their mid-gap spectral properties, and the manner in which these features can be controlled by altering the local phase profile. 

Our numerical technique is straightforward in discretizing the low-energy degrees of freedom given in Eq. \ref{Eq:Ham}. These Majorana fermion states are thus confined to a one-dimensional lattice having hard boundary conditions. The Hamiltonian can thus be represented in matrix form, consisting of the kinetic term and the coupling mediated by the spatially varying gap function. The eigenvalues obtained from diagonalizing the matrix thus correspond to discrete low energy states. We note here that for these Dirac-like models, discretization results in a "fermion doubling" issue \cite{Nielsen}; we are thus left with taking into account only half the eigenstates as a true representation of the spectrum. 

\begin{figure}

\subfloat[][]{\includegraphics[width=0.4\textwidth]{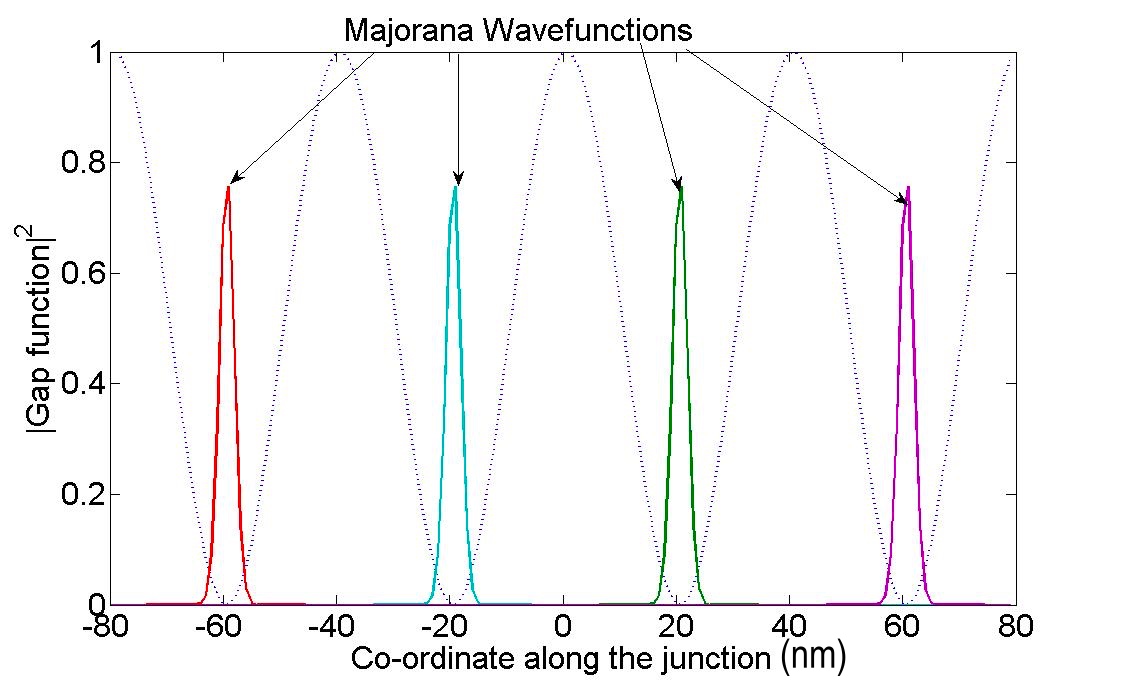}}\\
\subfloat[][]{\includegraphics[width=0.4\textwidth]{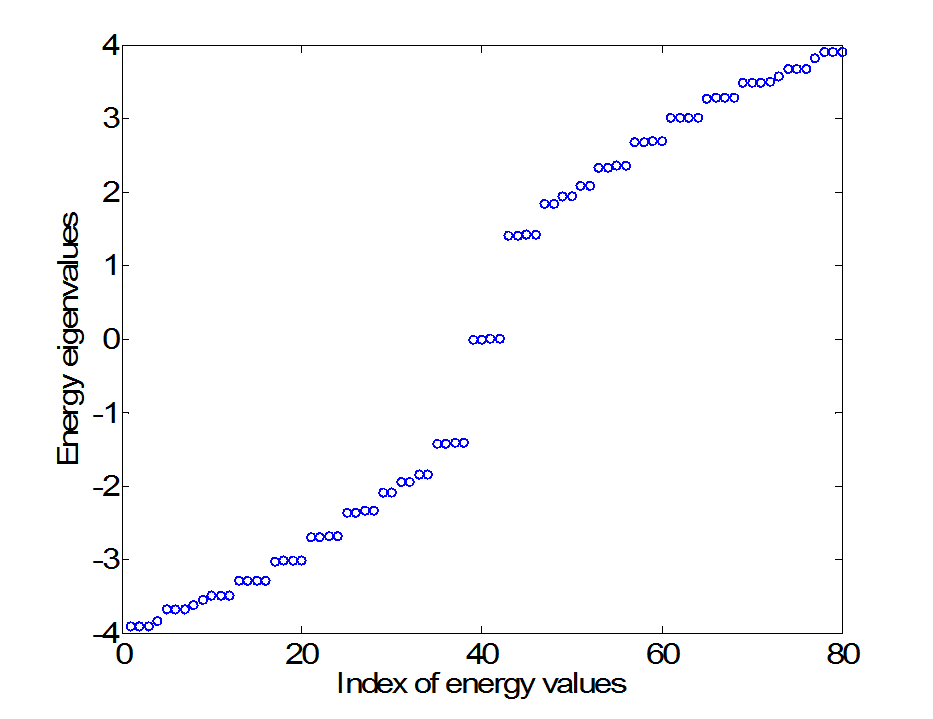}}\\
\caption{(a) The application of a magnetic field leads to variation of the phase difference along the Josephson junction and the gap function. The gap function, plotted as a function of distance along the junction, goes to zero when the SC phase difference crosses multiples of $\pi$.  Corresponding to every spatial location where the gap function goes to zero, there exists a localized Majorana mode - the corresponding wavefunction profiles are shown here. (b) The spectrum of Andreev bound states(in units of $\hbar v_M$) obtained from the diagonalisation of model Hamiltonian Eq. \ref{Eq:Ham} for the given gap function profile. The mid gap states correspond to the MBS.}
\label{numerics}
\end{figure}

  We first consider the instance where the phase variation in Eq. \ref{Eq:Ham} is linear $\phi=2\pi y/ \ell_B=2\pi y N/W$ as $\ell_B=W \Phi_0/\Phi_B=W/N$, where $N$ is the number of flux quanta. The phase therefore increases from $-\pi N$ to $\pi N$ as the coordinate along the junction, $y$, spans the junction from $-W/2$ to $W/2$ . This situation encompasses four  half-flux quanta within the junction, which ought to lead to four MBS. We explicitly ascertain this MBS distribution and related features by numerically diagonalizing the Hamiltonian in Eq. \ref{Eq:Ham}; Figure \ref{numerics} shows the numerical results. 

For this case of four flux quanta piercing the junction, Figure \ref{numerics} (a) shows the variation of the gap function along the junction. Correspondingly, Figure \ref{numerics}(b) shows the energy spectrum. Most energy states lie outside a gap region centered around zero energy. As expected, four states however are mid-gap states effectively at zero energy. Our analyses also show that with increasing flux, the formation of new MBSs occurs through select states lying outside the gap entering the gap region and nucleating towards zero energy. Plotting the eigenstates of the corresponding wavefunctions in Figure \ref{numerics}(b) indeed shows them to be isolated, evenly spaced, bound states localized along the junction at the zeroes of the gap function. Each of the bound states shows exponential decay in isolation. Moreover, the eigenfunction is completely real, making it of the Majorana form. The MBS wavefunction at a distance $\delta y$ away from its center respects the form
\beq
\label{eqn:gamma}
 \gamma(y) \approx f(x) e^{-|\delta y|/ \lambda_M}.
 \eeq 
Here, the decay length is characterized by $\lambda_M=\sqrt{\hbar v_M l_B/\Delta}$ and $f(x)$ describes the confinement of the MBS in the transverse direction.

The MBSs are effectively isolated when their separation is significantly greater than their decay length. However, when brought closer, a pair of MBSs becomes coupled due to the overlap in their wavefunctions \citep{Kitaev01}. 
This coupling between the neighboring MBSs, say $\gamma_a$ and $\gamma_b$ separated by distance $L_{ab}$, leads to an effective Hamiltonian of the tunneling form
\beq
H_{ab} \approx i t_{ab} \gamma_a \gamma_b, \quad  t_{ab} \approx  e^{-L_{ab}/\lambda_M}
\label{eq:coupling}
\eeq
This is due to the overlap of the wavefunctions of the two MBS  in the junction across the distance  $L_{ab}$ . This coupling results in a tunnel splitting between the degenerate zero energy states associated with the MBS pair. 

\begin{figure}
\subfloat{\includegraphics[width=0.45\textwidth]{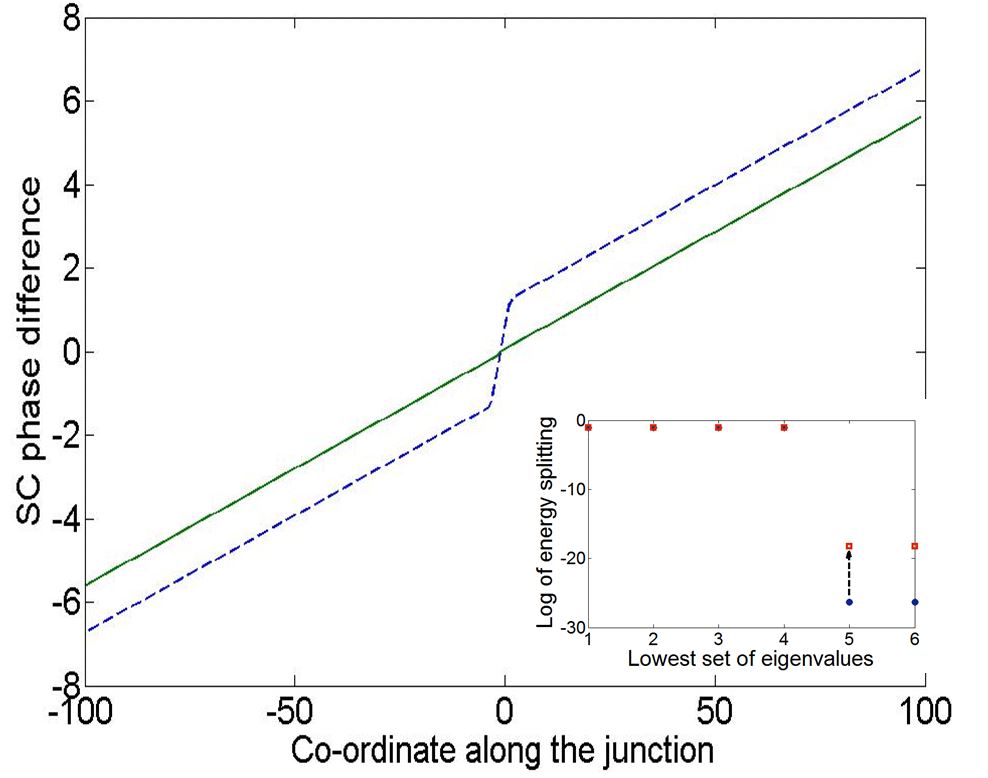}}\\
\subfloat[][]{\includegraphics[width=0.5\textwidth]{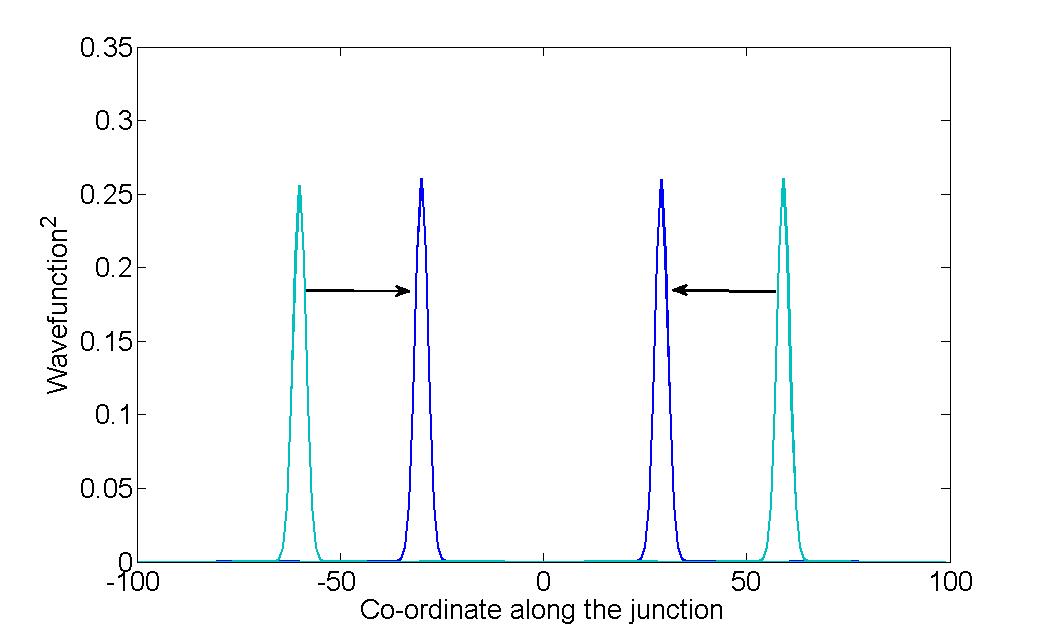}}
\caption{The phase profile with and without change in the local SC phase is shown in (a). The slope of the phase changes in a small region between the MBS. This results in a displacement of the MBS as shown in (b). There is a corresponding shift in the energy of MBS as shown in the inset of (a)}
\label{fig:MBStune}
\end{figure}

The proposed S-TI-S architecture here hinges on the ability to move and couple MBS pairs. Our proposed schemes for such manipulation primarily involve changing the local phase variation. As an example, consider the case of two MBSs initially far apart, as shown in Figure \ref{fig:MBStune}. Now, changing the local phase more rapidly between the two MBSs, as shown in Figure \ref{fig:MBStune}(a), decreases their separation, as shown in Figure \ref{fig:MBStune}(b). The inset in Figure \ref{fig:MBStune}(a) shows that the MBSs have come close enough to result in a numerically discernible tunnel splitting.

Such controlled MBS mobility and tunable coupling are essential ingredients in braiding schemes considered here. These schemes require the four-MBS configuration. Generalizing Eq. \ref{eq:coupling}, the tunnel coupled effective Hamiltonian takes the form
\beq
H= it_{12} \gamma_1 \gamma_2 +i t_{23} \gamma_2 \gamma_3 + it_{34} \gamma_3 \gamma_4
\eeq
We will see that this Hamiltonian can be used to demonstrate an effective braiding  by tuning one of the couplings.
In the case of semi-conductor wire heterostructures, the values of $t_{12}$, $t_{23}$, $t_{34}$ range around 0.5-30 $\mu eV$ \citep{Burnell14,Sarma12} . For S-TI-S junctions of order of $1\mu m$ width and MBS separate by $~0.1 \mu m$, the strength of coupling between them would be of the order of $10^{-1} \mu eV$. Thus, the estimates are comparable to those in the nanowire sutations and in both cases are highly sensitive to separation distance due to the exponential dependence.

\section{Josephson supercurrents and signatures of MBS states}
\label{sec:Josephson}

Having established the description of the S-TI-S Josephson junction, we are now in a position to derive the form of Josephson currents across the junction and the crucial role played by MBS contributions. In semiconductor nanowire systems, the zero-bias conductance peak in transport through the end of the wire is a signature of the presence of MBSs \cite{Das12,Churchill13,Rokhinson12,Deng12,Finck13,Mourik12}. Here, we propose that the onset of a 4$\pi$-periodic component in the Josephson current-phase relation, revealed by characteristic features in the critical current modulation patterns in a magnetic field, play a similar role in providing an indication of the presence of MBSs.

\subsection {Josephson Interferometry}

Josephson phase-coherent devices are the active elements in superconducting electronics and can be probed using various electrical circuit analysis methods. A Josephson junction consists of two superconducting islands separated by a barrier made of an insulator, normal metal, or, in our case, a topological insulator in which conductance through the topological surface states plays a key role. In a point junction, the defining features of these junctions are captured by specific relationships among the gauge-invariant phase difference $\phi$ across the junction, the voltage across the junction $V$, and the supercurrent though the junction $I_s$:  (i) the Josephson supercurrent $I_s$ is related to the phase via the current-phase relation. In an ordinary Josephson junction, $I_s=I_{c} \sin \phi$, where $I_{c}$, the critical current, is the maximum supercurrent that the junction can sustain, and (ii) a voltage causes a rate of change in phase given by $d\phi /dt = 2 \pi V/ \Phi_0$, where $\Phi_0$=h/2e is the flux quantum. The gauge-invariant phase difference between the superconductor electrodes is given by $\phi= \phi_1 -\phi_2-2 \pi/ \Phi_0 \int_1^2 {\bf A}\cdot d{\bf l} $, where, ${\bf A}$ is the vector potential associated with the magnetic field. The Josephson coupling free energy depends on the phase difference $F(\phi)=-I_{C0} (\Phi_0/2 \pi) \cos \phi$, and has a minimum energy at a phase difference of 0. The current-phase relation can be simply derived by taking a derivative of the free energy with respect to the phase \cite{Tinkham}. 

In an extended junction, a magnetic field can penetrate the junction barrier, creating a gradient in the phase difference along the width of the junction.  This gradient creates a corresponding variation of the supercurrent with position that depends on the local current-phase relation.  For a uniform magnetic field applied to a junction with a uniform critical current density and a conventional sinusoidal CPR, the maximum of the supercurrent varies with magnetic field according to a Fraunhofer pattern familiar from single-slit optical diffraction: $I_c(\Phi)= I _{c0} |\frac{\sin(\pi \Phi /\Phi_0) }{(\pi \Phi /\Phi_0)}|$, where $\Phi$ is the total flux passing through the junction. Spatial variations in the magnetic field, the Josephson critical current-density, the local order parameter symmetry, or the functional form of the CPR can all modify the shape of the critical current diffraction pattern, providing a way to characterize the properties of the junction.  This technique, collectively know as Josephson interferometry, has played an important role in determining the pairing symmetry of unconventional superconductors such as d-wave and p-wave superconductors. \citep{DaleRMP,Kidwingira} and in exploring supercurrent channels . In this paper, we use propose this technique as a means of revealing the nucleation of Majorana bound states via changes in the diffraction pattern.


The presence of the localized Majorana bound states (MBS) in the S-TI-S junctions make two distinct changes to the junction properties.  The first is that they add additional supercurrent channels localized near the locations in the junction where the phase is an odd-multiple of $\pi$, breaking the uniformity of the supercurrrent provided by the ordinary Cooper pair component. We note that while a single MBSs cannot transport charge in and of themselves, and would require the presence of other MBSs present, they define the regions through which charge transfer can occur across the junction. Second,the current-phase relation of these states adds a $4\pi$ periodic contribution to the CPR \citep{Kitaev01,Fu08}.  This arises because the MBS current is due to single-electron processes that lead to fractional Josephson phenomena\cite{Kwon04}, unlike the conventional Josephson junction processes that involve only tunneling of Cooper pairs across the junctions and hence $2e$-processes. The CPR thus acquires an additional current contribution $I_{Ms}=I_{Mc} \sin(\phi/2)$. Here, based on the model established in Sec. \ref{sec:model},  we evaluate the manner in which MBS states contribute to the supercurrent and modify the critical current diffraction pattern in a characteristic way.

\subsection {Derivation of MBS contribution to Josephson current}

The spatial dependence of the MBS contribution to the supercurrent is calculated from the location and magnitude of the Majorana wavefunction. Note that although these act like local inhomogeneities in the supercurrent density distribution in the junction, they are located where the local phase difference in the junction is an odd-multiple of $\pi$ rather than at discrete fixed defects in the junctions.  They thus are required to move in spacing and position as the magnetic field or current applied to the junction are changed.  It also means that to determine the positions of the MBS and the diffraction pattern at every field or current, we need to first find the phase at one location in the junction that minimizes the S-TI-S junction energy, which in turn allows us to calculate the total supercurrent and its spatial variation.  In the short junction limit for which the fields generated by the Josephson currents are negligible compared to the applied magnetic field, the dominant energy is the Josephson coupling energy; we therefore maximize the total supercurrent. Given the complex spatial landscape and associated inhomogeneous current distribution, adapting the point junction approach of relating the supercurrent to a derivative in the free energy can be convoluted; our phenomenological approach thus directly applies input on the MBS spatial profile from the previous section to the expected local form of the supercurrent as a function of the phase $\phi$. While details of the supercurrent behavior would depend on the exact system and geometry, here we pinpoint the universal features emerging from the presence of spatially distributed MBSs.

The local supercurrent distribution includes the Cooper pair current, which we assume is uniform along the junction width and has a conventional $\sin(\phi)$-dependence, and the localized Majorana bound state currents which have a $\sin(\phi/2)$-dependence.  In general, there may also be contributions from high-transparency low-energy surface states in locations where the energy gap is small.  These give rise to a skewed supercurrent contribution to the  CPR that has higher-order harmonic components \cite{Titov,Manjarres,SkewSochnikov}. These have been extensively studied in graphene-barrier Josephson junctions.  In this treatment, we neglect them because they do not contribute to the lifting of nodes in the diffraction pattern that is the signature of the MBS on which we will focus.  

For a junction of of width $W$, we use the phase in the center of the junction $\phi_0$ as the reference phase. Including both the Cooper pair and Majorana bound state contributions to the current, the local supercurrent as a function of y, the position along the junction, is then 
\beq
\label{eqn:jy}
J(y)=J^{CP}(y) \sin(\phi(y))+J^{MBS}(y)\sin(\phi(y)/2), 
\eeq
where the local phase difference at y is given by 
\beq
\phi(y)=2\pi (\Phi/\Phi_0) (y/W) + \phi_0.
\eeq

\begin{figure}
\includegraphics[width=0.95
\columnwidth]{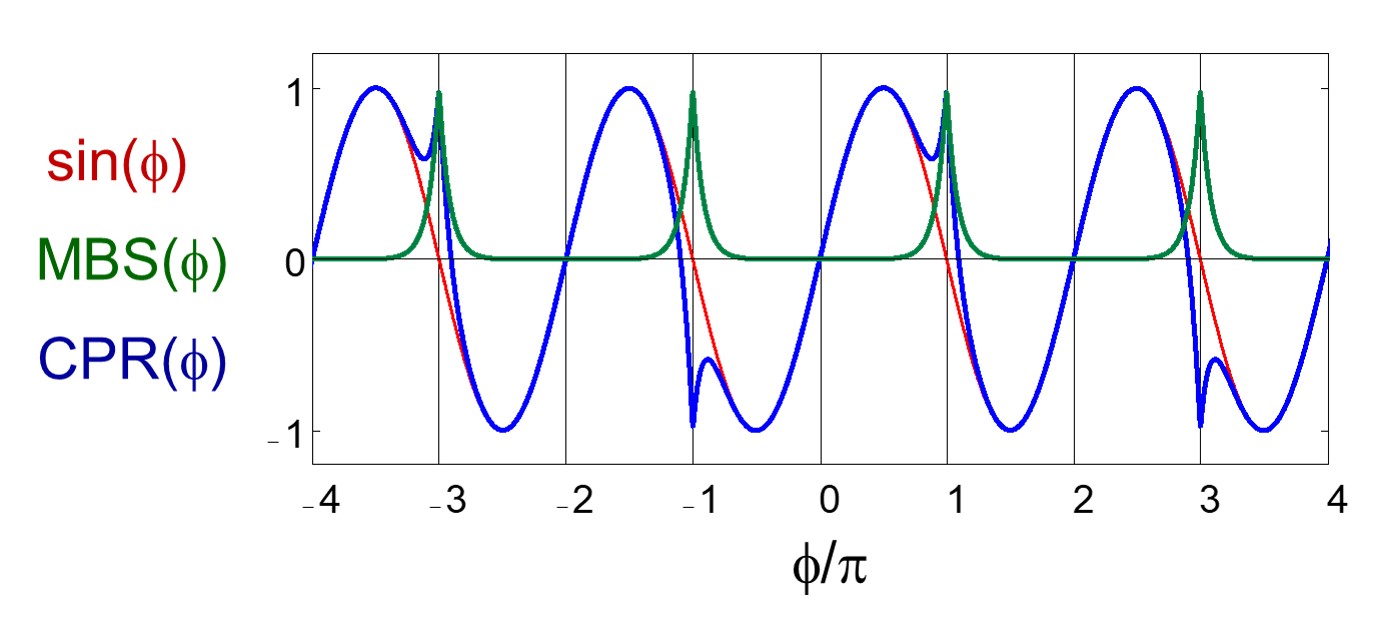}
\caption{Local supercurrent vs. local phase difference for the Cooper pair current $J^{CP}(\phi)$ (red), the Majorana bound state current $J^{MBS}(\phi)$ (green), and the total combined current $J^{CPR}(\phi)$ (blue).}
\label{cpr}
\end{figure}

Components of Eq. \ref{eqn:jy} are plotted in Figure \ref{cpr}. We assume here that $J^{CP}(y)$ is uniform across the junction and that $J^{MBS}(y)$ is localized on the Josephson vortices where the local phase difference is an odd multiple of $\pi$ and given by Eq. \ref{eqn:gamma} for each of the MBS in the junction.  Here, with regards to the form of the Majorana current, we make the assumption that charge $e$ transfer associated with the $\sin(\phi(y)/2)$ term only takes place through the MBS, even if via spatially separated pairs. Thus, mirroring the form of the localized MBS wavefunctions, we phenomenologically model the MBS current as having the same spatial distribution and also to be consistent with recent experiments\cite{ExperimentPaper,Ghatak18}. The distribution of J(y) is chosen phenomenologically as a localised distribution, such as exponential and gaussian distributions in y which integrates to 1. To find the supercurrent, we integrate across the junction and maximize the current as a function of the phase at the center of the junction.

The result of this calculation depends on the relative sizes of the Cooper pair and Majorana currents and the details of the Majorana states that determine the width of the MBS wavefunctions. However, independent of the details, the signature feature of Majorana physics is the $4\pi$-periodic $\sin(\phi/2)$ contribution to the current which lifts only the odd nodes in the Fraunhofer diffraction pattern. This behaviour can ultimately be traced back to the $\cos(\phi/2)$ variation of $\phi$ in Eq. \ref{Eq:Ham}.

While this approach is able to capture the key feature of dispersive Majorana modes and MBS physics, experimental settings can add additional complexities. The current density in the junction can be inhomogeneous and the magnetic field can be non-uniform as a result of screening effects or trapped vortices.  The current-phase relation of the junction can also be more complex \cite{Likharev}  due to various factors such as the nature of the non-superconducting region, unconventional pairing symmetries, dispersive modes, and scattering.

\subsection {Signatures of Majorana bound states in diffraction patterns}

Critical current diffraction patterns measure the maximum supercurrent of Josephson devices as a function of applied magnetic field and are one of the most powerful methods for characterizing their properties. The presence of magnetic flux threading the junction barrier creates a continuous variation of the superconducting phase difference across the width of the junction. This results in interference of the supercurrents at different points along the junction having different phases and leads to a diffraction pattern.  Diffraction patterns that deviate from the Fraunhofer form $I _{C0} |\frac{\sin(\pi \Phi /\Phi_0) }{(\pi \Phi /\Phi_0)}|$ familiar from single-slit optical interference can reveal valuable information about the junction structure and the mechanisms that carry supercurrent. In our case, we will use this metric to probe the MBSs and the associated single-electron tunneling processes they enable. 

\begin{figure}
\subfloat[][]{\includegraphics[width=0.95\columnwidth]{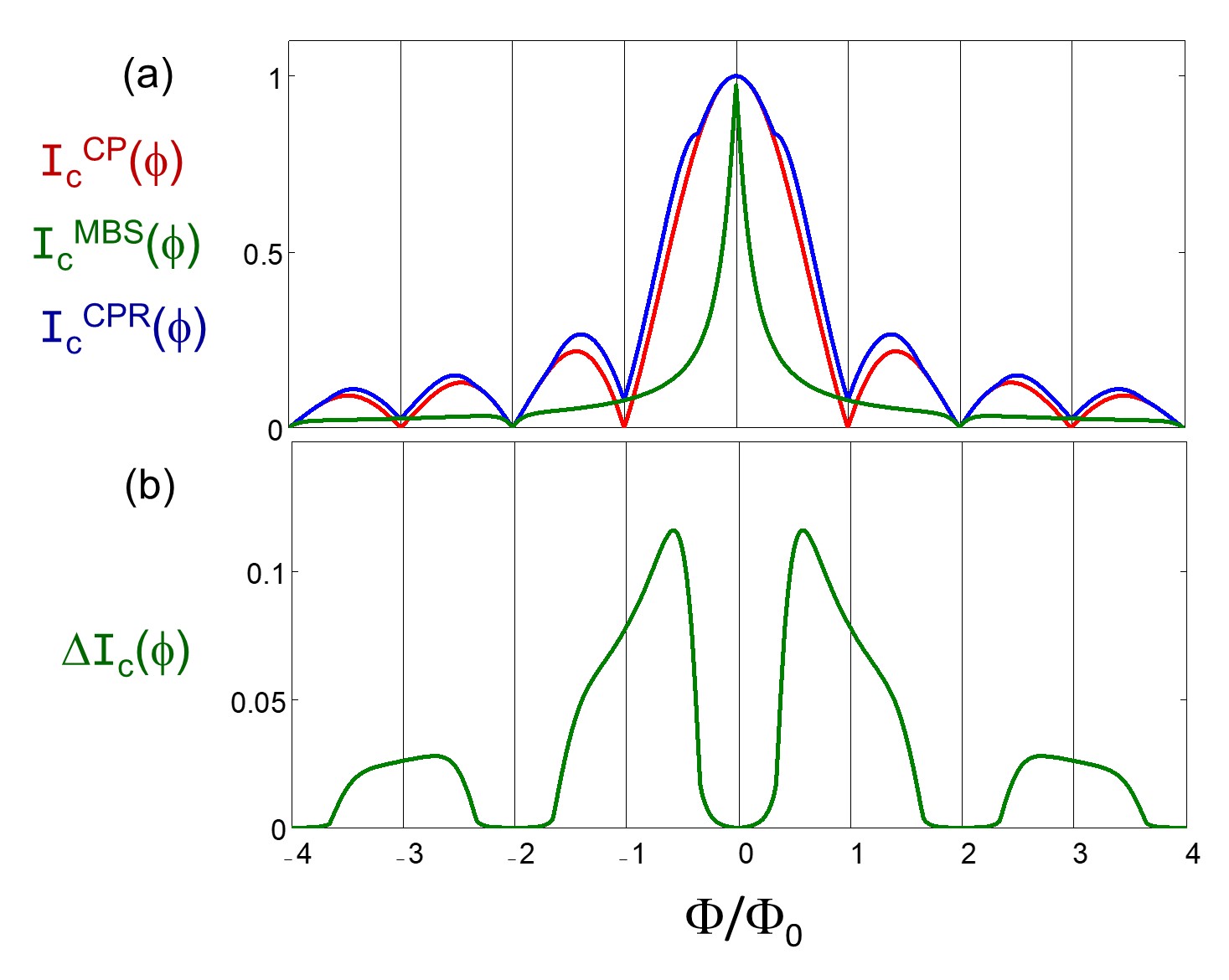}}\\
\caption{(a) The Josephson supercurrent vs. applied magnetic flux in the junction for only the Cooper pair current $I_c^{CP}$ (red), only the Majorana bound state current $I_c^{MBS}$ (green), and the total combined current $I_c^{CPR}$ (blue).  (b) The difference between the supercurrent with and without the Majorana bound state contributions.}
\label{diffraction}
\end{figure}

We employ the procedure outlined above to numerically evaluate the diffraction patterns in the S-TI-S junction; the results are shown in Figure \ref{diffraction}. As the flux through the entire junction is varied, the number of Josephson vortices in the junction changes as does the corresponding number of MBS pairs. In addition to the regular Cooper pair contribution, the MBS contribute to the supercurrent. Figure \ref{diffraction}(a) shows the separate contributions of a uniform Cooper pair current $I_c^{CP}$ (red) and the localized Majorana bound states $I_c^{MBS}$ (green) if they were the only contribution to the supercurrent. The distinct difference in the field spacing of the nodes in these two components shows the role of the $\sin(\phi)$ and $\sin(\phi/2)$ dependences of the two supercurrent terms, reflecting the regular Cooper pair processes (with flux period $h/2e$) vs MBS mediated single electron processes (with flux period $h/e$) of the Majorana states. 

The total critical current diffraction pattern $I_c^{CPR}$ (blue) shows the critical current calculated for the combined Cooper pair and Majorana processes.  Note that this is not just the sum of the two components because of the interference of the two terms in the maximization of the critical current described above.  The key feature is the lifting of the odd-valued nodes. This feature is a characteristic signature of MBSs localized within the junction.  There are also additional features that occur at magnetic fields at which Majorana bound states enter the junction.  These can be more easily identified by looking at the difference between the critical current diffraction pattern with the MBSs included and the Fraunhofer pattern expected in an ordinary Josphson junction, shown in Figure \ref{diffraction}(b). As expected, the Majorana contribution enhances the supercurrent around the odd-numbered nodes in the Fraunhofer diffration pattern, creating the lifting of those nodes that is a signature of the 4$\pi$-periodicity of the Majorana states.



We also note that the results shown in Figure \ref{diffraction} depend on the parameters of the system, in particular the relative size of the Cooper pair and Majorana contributions of the supercurrents and the extent of localization of the MBS. Those are expected to depend on junction and interface properties and could widely vary between different experimental settings. Nevertheless, as long as the Majorana contribution is even a small discernible fraction, the node lifting feature would be measurable. Experiments have indeed observed such features \citep{Sochnikov13,ExperimentPaper} and allow determination of those details.

In the S-TI-S junctions, the node-lifting of the odd nodes in the magnetic diffraction patterns is perhaps the most obvious signatures of the Majorana states, equivalent in some ways to the zero-bias conductance peaks that have been interpreted evidence for Majorana modes in proximized semiconductor nanowires. In the nanowire system, many subtleties in interpreting the zero-bias peaks as a definitive signature of MBS have been identified, e.g. other bound states due to disorder and resonances that contribute to single electron processes could lead to spurious effects. Similarly, in our system of S-TI-S junctions such single electron processes could provide a challenge in definitively pinpointing the contribution of MBS to the critical current diffraction patterns. Because of these complicating factors, it is fair to say that, in common with the nanowire system, only the demonstration of braiding and associated parity switches will constitute definitive evidence for Majorana fermion physics.

\section{Non-Abelian rotation in the ground state manifold of Majorana modes }
\label{sec:MZM}
The primary building blocks for non-Abelian rotations and braiding within the Josephson junction architecture are contained in Figure \ref{fig:1}. The elements forming the basis of these non-Abelian rotation are the MBSs localized at Josephson vortices. Here we briefly outline the underlying principles, discussing the relevant Hilbert space, operations, and the physical manifestations associated with these rotations. 

\subsection{Braiding through physical exchange of the MBSs}
The simplest instance of braiding through involves four MBSs, say denoted by   $\gamma_1$, $\gamma_2$, $\gamma_3$, $\gamma_4$. Any pair of MBSs forms a Dirac (electronic) state that can be occupied or not. As a specific choice, consider $c_A= (\gamma_1+i \gamma_2)/2$ ,$c_B=(\gamma_3+i \gamma_4)/2$. In the absence of coupling between the MBSs, the degenerate ground state manifold is spanned by 
\beq 
\ket{N_A,N_B}:\ket{0,0},\ket{1,1}, \ket{1,0}, \ket{0,1}
\eeq
 where $N_A, N_B$ denote the occupation of the electronic states. For N pairs of MBSs, the ground state is $2^N$ fold degenerate. The occupation of all such parity states decides the net fermionic parity of the ground of the system. Thus unlike conventional superconducting ground state, which is always a superposition of states having an even number of electrons in form of Cooper pairs, a topological superconductor can have states with net fermion parity to be either even or odd.

 The simplest braiding operation is an exchange in the positions of the two MBSs in such a way that the underlying ground state degeneracy remains. This is a topologically robust operation. How does this exchange in the position space affect the space of ground state? It can be shown  that \citep{Ivanov01} the exchange of two Majoranas $\gamma_i$,$\gamma_j$ is represented in the ground state manifold as a unitary rotation in the space $\{\ket{0,0},\ket{1,1}, \ket{1,0}, \ket{0,1}\}$ given by 
 \beq
 U_{ij}= exp(\pm  \pi \gamma_i \gamma_j/4). 
\eeq 
  For example, if we start with a state $\ket{0,0}$, then exchanging $\gamma_2, \gamma_3$ results in 
  \beq
  U_{23} \ket{0,0}= (\ket{0,0}-i \ket{1,1})/ \sqrt{2}
  \eeq
 In principle, one can track such rotations by measuring the fermion occupation i.e the fermion parity of the electronic states in the ground-state manifold. The order of consecutive exchanges matter as the unitary operations do not commute: $U_{12}U_{23} \neq U_{23}U_{12}$. Thus the name non-Abelian rotation.
  
 The actual implementation of such rotations in our proposed architecture involves sequences of vortex motion such as shown in the trijunction geometry of Figure \ref{fig:3}. In the next section, we provide the exact experimental steps to perform these sequences. We remark here that these sequences are in the spirit of the original proposal by Ivanov \cite{Ivanov01} for performing exchange operations.

\subsection{Effective braiding through tuning MBS hybridization}
A key feature of the topological qubit formed by the electron parity state is its non-locality. It is shared by two MBS states confined to vortices that can be very far apart.  We have seen the manner in which physical exchange results in non-Abelian rotations in the Hilbert space of these parity states. An alternate method for performing non-Abelian rotations without physical exchange involves tuning the coupling between an MBS pair and temporarily splitting the associated degeneracy through hybridization.

As a specific example, consider four MBSs $(\gamma_1,\gamma_{2},\gamma_3,\gamma_{4})$, this time with their vortex cores aligned along a junction, as in Figure \ref{fig:3}(b). The effective low energy Hamiltonian of this system, as discussed in Sec. \ref{sec:model}, is given by
\beq
H_{12}=it_{12}\gamma_1 \gamma_{2}+ it_{23}\gamma_{2} \gamma_3 + it_{34} \gamma_3 \gamma_{4}. 
\label{eq:H12}
\eeq
Here, MBSs $\gamma_{1},\gamma_2$ are coupled with strength $t_{12}$, MBSs $\gamma_{2}\gamma_3$ with strength $t_{23}$, and $\gamma_{3},\gamma_4$ with $t_{34}$ as shown in Figure \ref{fig:3}. Now let us denote the non-local electronic states by $\Gamma_1 = (\gamma_1 + i \gamma_2)/\sqrt{2}$ and $\Gamma_2 = (\gamma_3 + i \gamma_4)/\sqrt{2}$. The occupation of these modes is given by $N_1= \Gamma_1^{\dagger} \Gamma_1$ and $N_2 = \Gamma_2^{\dagger} \Gamma_2$. As with the exchange braiding case, the Hilbert space of the system is given by the occupation of these 2 states $\ket{N_1,N_2}$ : $\ket{0,0},\ket{1,1}, \ket{1,0}, \ket{0,1}$. The Hamiltonian in this Hilbert space is then block diagonal. We focus only on even parity block corresponding to $\ket{0,0},\ket{1,1}$; the odd parity block is decoupled and contains analogous physics.
In the reduced two-component basis of even parity states, the tunnel coupled Hamiltonian of Eq.\ref{eq:H12} takes the form
\beq
H_{e12} = \left( \begin{array}{cc}
t_{12}+t_{34} &  t_{23} \\
t_{23} & -(t_{12}+t_{34}) \end{array} \right)
\eeq
Treating the two states of the Hilbert space $\ket{0,0},\ket{1,1}$ as the "spin-up" and "spin-down" eigenstates of Pauli matrix $\sigma_z$ respectively, we can cast the Hamiltonian in terms of Pauli matrices  as:
\beq
H_{e12}= (t_{12}+t_{34})\sigma_z + t_{23} \sigma_x
\eeq
 Preparing the system in an initial state, say "spin up" $\ket{0,0}$  and then changing the $t_{23}$ ("a transverse field") would result in the rotation of the state in the spin basis. Effectively, changing the coupling in two Majorana modes $\gamma_2,\gamma_3$ would induce non-local parity correlations.  It has been explicitly  shown that these rotations are equivalent to non-Abelian rotations in the ground state manifold. The detailed proof of this is equivalence is much involved and we refer the reader to \citep{Burnell14,Chiu15,Sau11} for a detailed treatment. These works show that though such a protocol would not lead to a full-fledged non-Abelian braiding, it nevertheless indicate the entanglement characteristics of non-Abelian statistics. 
 
Specific sequences of such effective braiding would involve preparing the system in a prescribed initial state in the degenerate Hilbert space, bringing a pair of MBSs to break the degeneracy via coupling, and time evolving the initial state in a manner prescribed by Eq. \ref{eq:H12}. The time scale for varying the coupling is set by the maximum degeneracy splitting; compared to actual braiding, which involves the robust topological operation of exchange, this time scale dependence poses a limitation. Nevertheless, given enough experimental control and knowledge of tuning parameters, qubit operations can be made viable through this procedure.

The operations involving such hybridization can only be performed with adjacent pairs in the case of a single extended junction. As a first step, such operations would be important in determining degeneracy splitting and coherent times. Pairwise operations could also progress between adjacent MBSs along the junction. However, the full-fledged set of allowed operations would require more complex Y-junction geometries.

\section{Implementation of schemes for braiding and hybridization}

We have  proposed two schemes for braiding MBSs, both based on the ability to manipulate MBSs by moving the Josephson vortices to which they are bound in networks of S-TI-S junctions.  This can be done easily and controllably by applying a combination of magnetic fields and phase biases (via currents) to the junctions.  For a uniform junction structure, in which the separation of the superconducting electrodes is the same along the width of the junction, in a uniform magnetic field perpendicular to the direction of the supercurrent, the separation of MBSs/vortices is set only by the magnitude of the field, with one quantum of flux threading the junction between each adjacent pair of MBSs.  This assumes that supercurrents in the junction do not generate significant magnetic fields, the so-called “short junction” limit which is satisfied in the proposed devices because of the small magnitude of their supercurrent densities.  Figure \ref{fig:2}(b) shows the location of vortices/MBSs in the junction as a function of field --- with no currents in the junction the vortices enter the junction symmetrically from the edges as the field is increased.  The location of the vortex chain within the junction can be shifted by applying a current through the junction that induces a phase drop across the junction.  If the critical current is exceeded, a voltage is induced across the junction. This causes the vortices to move across the junction as the phase winds according to the Josephson relation.  This provides a way to move the MBSs (at very fast speeds $>1$km/s), but it also generates quasiparticles that could cause undesirable parity transitions.

\subsection{Non-Abelian rotations by exchange}

\begin{figure}
\includegraphics[width=1.0\columnwidth]{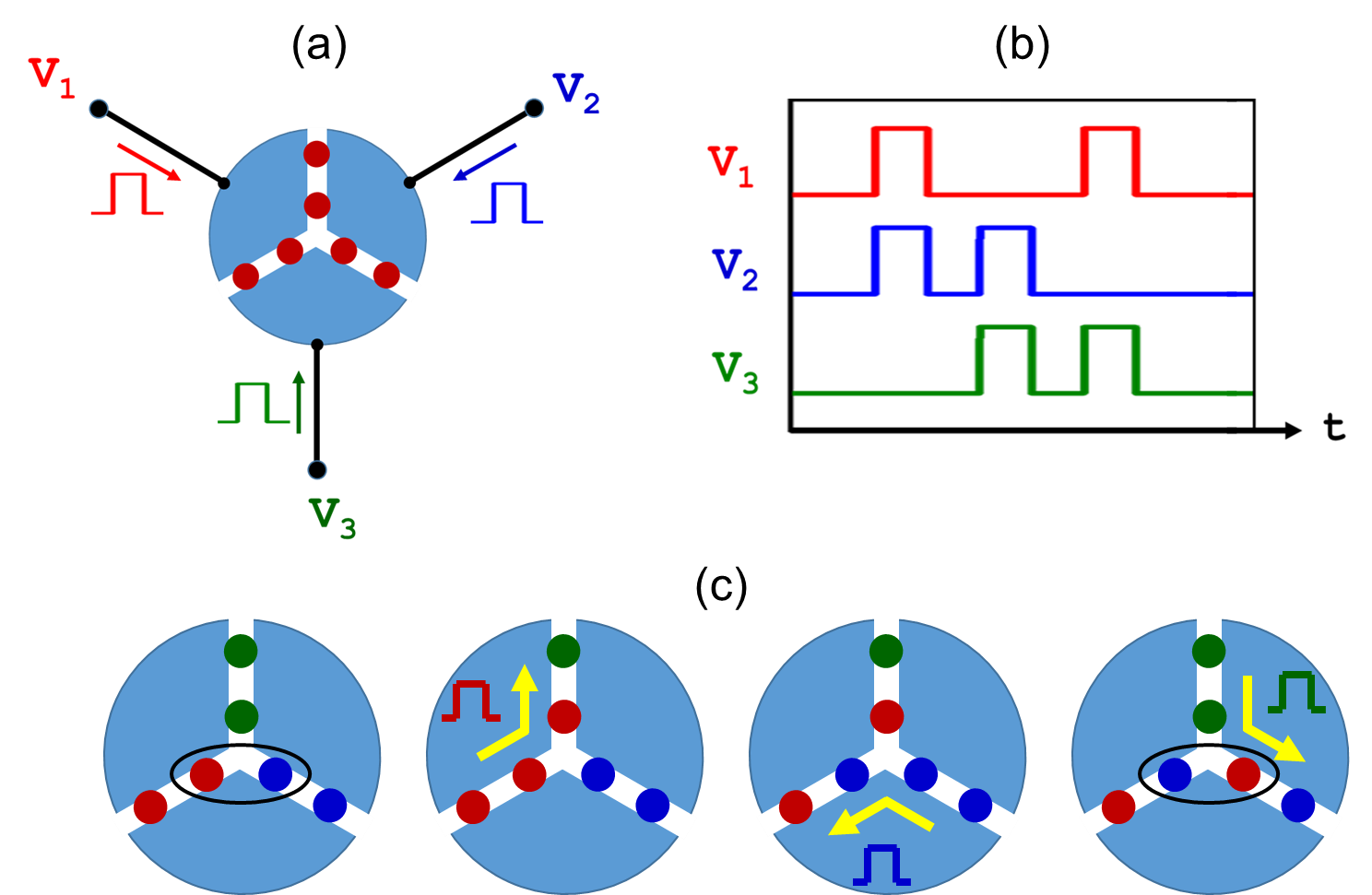}
\caption{Voltage pulse braiding: (a) Trijunction configured to allow application of Rapid Single Flux Quantum (RSFQ) pulses to the three superconducting electrodes. (b) RSFQ pulse sequence required to effect the exchange of Majorana vortices shown in (c).} 
\label{fig:v_braiding}
\end{figure}

In a trijunction consisting of three Josephson-coupled superconducting islands on a topological insulator film, as in Figure \ref{fig:v_braiding}, MBSs nucleate at the locations where the phase difference is an odd multiple of $\pi$, i.e. at the “cores” of Josephson vortices.  As we have described above, for a uniform structure in a uniform applied magnetic field, the separation of MBSs/Josephson vortices is set only by the applied field, and the vortex pattern is symmetric in the absence of applied currents.  However, this pattern can be manipulated by adjusting the relative phases of the islands, changing the location of the chain of vortices in each segment of the trijunction. 

One approach is to apply a sequence of Rapid Single Flux Quantum (RSFQ)\cite{Likharev_1991} voltage pulses across the junctions. Such pulses apply an integrated flux of one flux quantum, changing the phase across the junction by 2$\pi$ and shifting the vortices by one vortex spacing. They can be an be generated by Josephson junction circuitry developed for superconducting logic technology\cite{LIKHAREV_2012,Liebermann16, Leonard19}.

 In Figure \ref{fig:v_braiding}, we show a sequence of RSFQ pulses that effects an exchange braiding operation of two MBSs. In junctions in which the I$_c$R-product is of order 1$\mu$V, typical RSFQ pulses are of duration of order 1ns, so braiding operations can be accomplished at GHz frequencies. The disadvantage is that the finite voltage pulses will create heating and generate quasiparticle excitations that can cause quasiparticle poisoning which limits the parity lifetimes, degrading speed and performance.

\begin{figure}
\includegraphics[width=1.0\columnwidth]{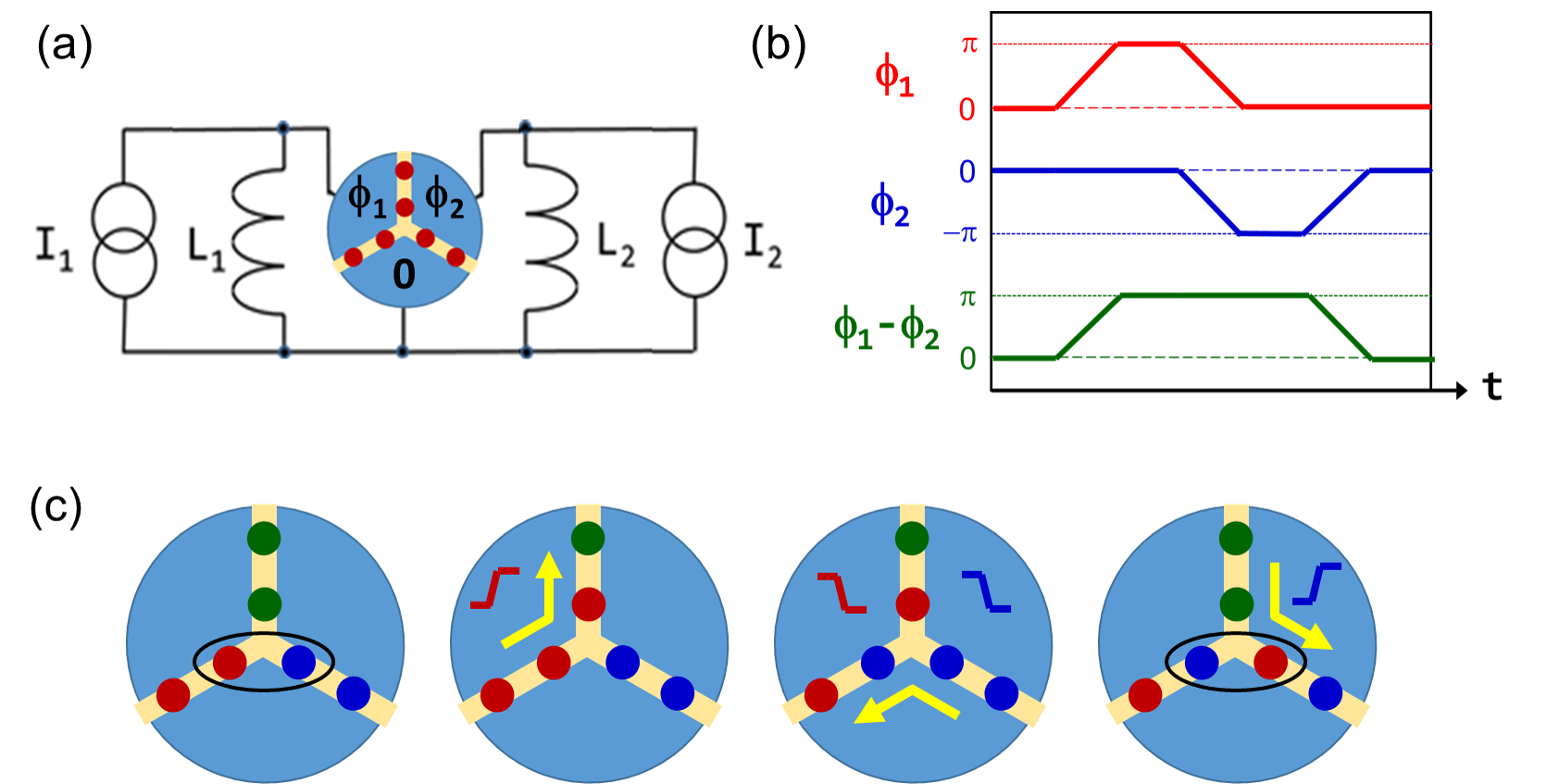}
\caption{Phase pulse braiding: (a) Trijunction configured to allow manipulation of Josephson vortices via phase control -- this requires superconducting inductances in parallel with two of the junction arms. (b) Pulse sequence of phase differences across the three junction arms required to carry out the braiding protocol shown in (c).  For small inductances such that most of the current flows through the inductors, the phase differences are nearly proportional to the applied currents $I_1$ and $I_2$.}\label{fig:10}
\end{figure}

A better technique can be achieved by shorting two of the junctions with superconducting loops and coupling flux into the loops via applied currents to induce phase differences, as in Figure \ref{fig:10}(a).  Note that the multiply-connected geometry allows us to access all possible phases across the junction so that we have complete control of the vortex locations without exceeding the critical currents and generating quasiparticle excitations.

In the proposed system, we can use this scheme to braid (exchange) pairs of MBSs, as illustrated in Figure \ref{fig:10}. Winding the relative phase of island 1 by $2\pi$ shifts all of vortices in the top and left junctions by one vortex spacing; repeating this for islands 0 and then 2 returns a vortex pattern with two vortices exchanged.  

As described in previous sections, in both schemes, the MBSs in each segment of the junction form parity qubits; the exchange operation corresponds to a braiding rotation in the basis of these qubit states, which should result in a change in the parity of the Majorana pair. The principle behind this operation is the same as that proposed for braiding MBSs in semiconducting nanowires in the T-junction geometry \citep{Alicea11}.

 \subsection{Non-Abelian rotation by hybridization}

\begin{figure}
\includegraphics[width=1.0\columnwidth]{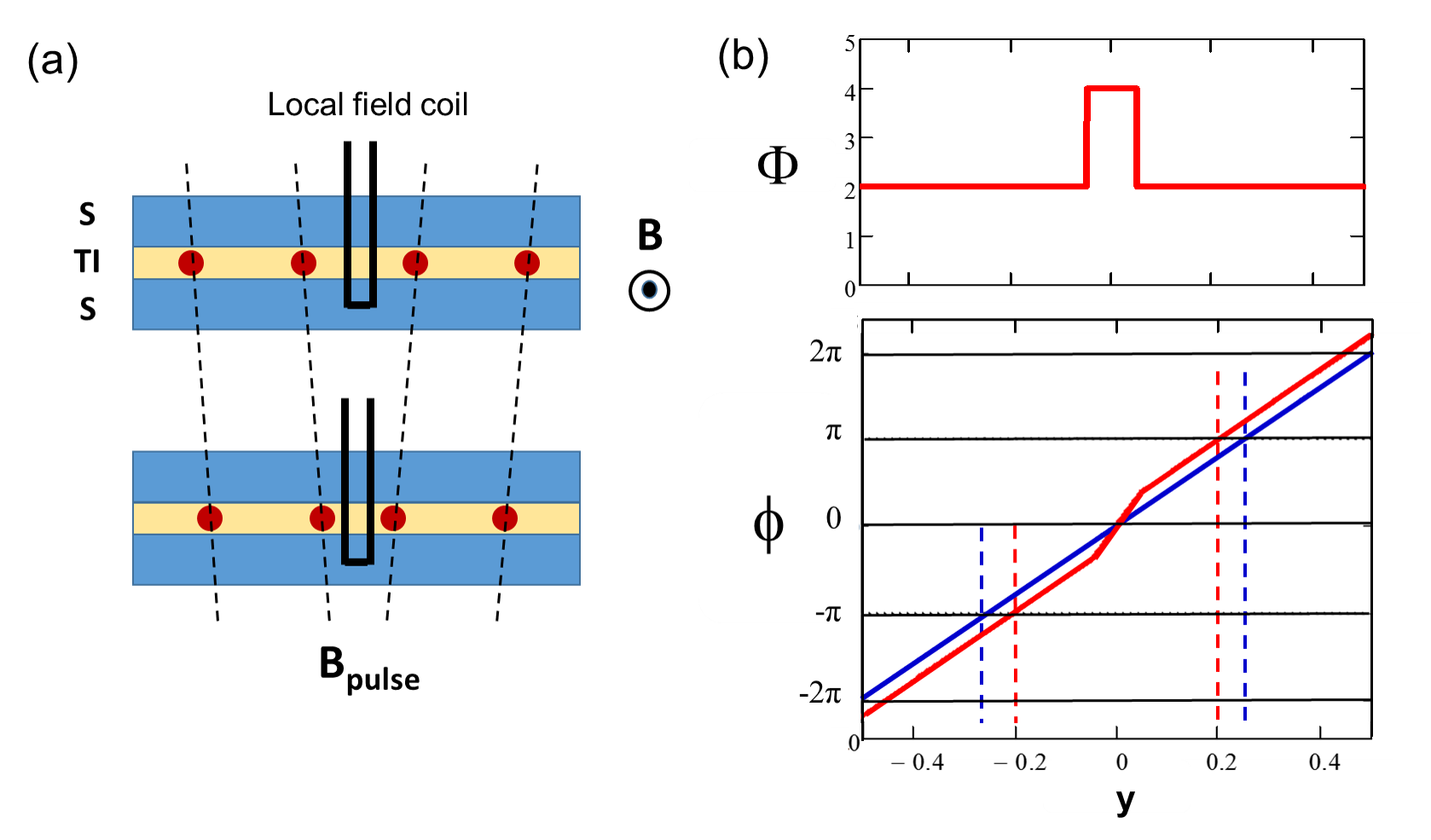}
\caption{Hybridization scheme:  (a) A current loop fabricated across the junction is used to (b) inject a magnetic field pulse, bringing the vortices closer or farther apart.}\label{fig:jun}
\end{figure}

As also described in previous section,we can alternatively carry out non-Abelian operations in a single S-TI-S Josephson junction, by bringing them close together resulting in hybridization and level splitting. As a physical implementation of such a scheme, an applied field creates a row of Josephson vortices evenly spaced by one flux quantum threading the junction, as in Figure \ref{fig:jun}.  A current pulse applied to a narrow loop of wire crossing the junction generates a localized increase in the magnetic field in a region between the vortices, bringing the vortices closer together.  This in turn changes the distance between the MBSs, creating an overlap of the wavefunctions of the MBSs and inducing a parity flip through hybridization.  Such controlled dynamic coupling constitutes the realization of an alternate scheme. The related non-Abelian operation depend on the magnitude of the energy level splitting in their time evolution. They thus are not topologically-protected, in contrast to exchange operations. Instead, the state evolution results in a parity transition whose dynamics depends on the exact protocol for the hybridization of the Majorana modes. A detailed analysis and realistic simulation of system with magnetic pulse dynamics and state evolution deserves additional attention.

\section{Implementation of parity readout schemes}

A key measurable consequence of MBS based non-Abelian operations is the associated non-local fermion parity transitions.
To demonstrate the occurrence of those transitions requires developing fast and high fidelity schemes to read the parity associated with specific MBS pairs.  Although that is not the primary subject of the work presented here, we will comment on how multiple viable schemes for achieving this.  These include:  \smallskip 

(1) Measurements of the critical current distribution of S-TI-S Josephson junctions performed by ramping the applied current and recording the value at the onset of a finite voltage across the junction.  The supercurrent has two contributions, one from the Cooper pair and one from the MBS. The parity states of MBS pairs results in the splitting of the critical current distribution into a bimodal distribution \cite{Woerkom15}. This approach is straightforward but is an intrinsically dissipative process, potentially contributing to quasiparticle generation and resulting parity transitions, so-called quasiparticle poisoning.  \smallskip 

(2) Coupling the junction to a microwave resonator and detecting the shift in the resonant frequency that depends on the energy-splitting from the Majorana fermion parity states.  This non-dissipative readout scheme is the basis of the measurements of transmon superconducting qubits \cite{Ginossar14}.  Microwave induced-qubit transitions involve a basis of states composed of multiple Cooper pairs. Here the MBSs offer the unique situation of an additional spectrum of states corresponding to odd-multiple of single electronic states.   \smallskip 

(3) Coupling the MBS pair to a single-electron transistor (SET) and measuring the conductance shift that is sensitive to the associated parity state \cite{Ben-Shach15}. \smallskip 

\begin{figure}
\includegraphics[width=1.0\columnwidth]{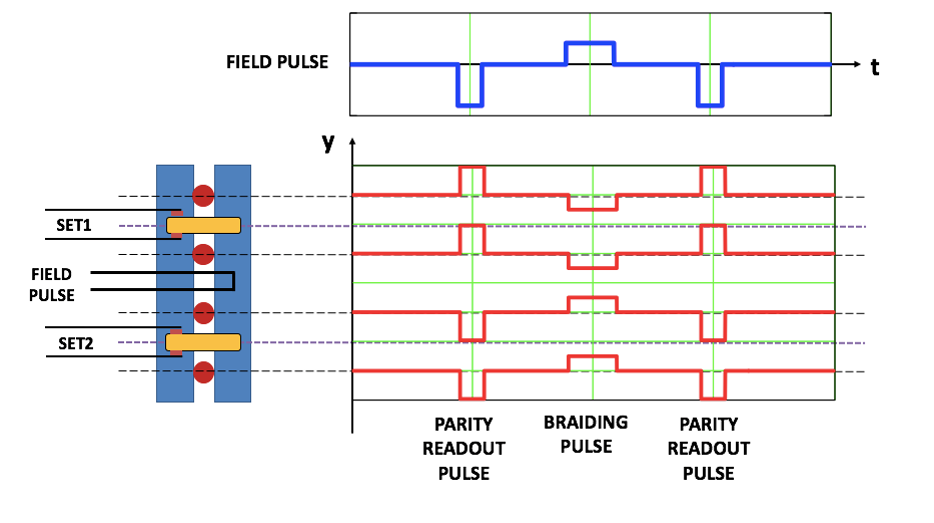}
\caption{Design of an experiment to demonstrate a parity exchange induced by braiding via a sequence of three magnetic field pulses:  (1) a negative magnetic field moves the MBSs to a positions underneath two SET detectors so that their parity can be read, (2) a positive pulse bring them together for a specific time, affecting a hybridization braid, and (3) a second negative pulse reads the resulting parity to test for a parity change.} \label{fig:11}
\end{figure}  

In Figure \ref{fig:11}, we show an example of a test circuit and pulse sequence that we propose to demonstrate parity changes via hybridization.  In this scheme, a negative magnetic field pulse applied to a coil in the center of the junction shifts the two middle vortices apart, moving them underneath the SET detectors to read their parity and thus prepare the initial qubit state.  This is followed by a positive pulse that brings the vortices together for a specific time required to induce a parity change.  This is followed by a second negative readout  pulse. Similar circuits can be used to test for parity changes following exchange braiding operations. 

This scheme provides a way to tune the coupling of the MBSs to the quantum dots for achieving fast and efficient readout.  It can be used to demonstrate parity changes and to determine parity lifetimes that limit MBS qubit performance. This will enable optimization of the distribution and manipulation of vortices and the pulse timescales and sequences required to perform MBS braiding operations and their effect on their parity.

Quasi-particle poisoning is one of the serious roadblocks in realizing braiding and non-Abelian rotation. The key measurement that indicates non-Abelian rotation is the correlated parity measurement of the non-local Majorana pairs. Spurious single-particle excitations due to quasiparticle poisoning in superconductors would make it hard to discern parity transitions due to non-Abelian rotation. Addressing this has been an open long-standing question under investigation even in the case of the extensively studies nanowires \cite{Rainis12,Karzig17,Karzig20}, and warrants a similar study in this  S-TI-S junction case.

 These different schemes are far reaching. Learning from the example of the nano-wire system, even the first step of detecting the zero-bias peaks makes for a landmark experimental test of Majorana physics. Here too, successful implementation of any one of these schemes would constitute a significant step.

\section{Status of experiments on the S-TI-S system}
Our goal in this paper was to present a generic description of the S-TI-S lateral Josephson junction platform and how it might be implemented for topologically-protected braiding operations.  In parallel, several experimental groups have begun to actively study S-TI-S junctions for prospective MBS realizations and manipulations, including the experimental contributors to this this work \cite{Sochnikov13,Ren19,Kurter14,Kurter15,Veldhorst_2012}.  Our calculations make specific testable predictions of a number of signatures consistent with the formation of topological surface states and Majorana bound states that should be revealed in measurements of the transport and Josephson properties of S-TI-S junctions. 

The first most significant observation would be a lifting of the odd-numbered nodes in the critical current, evidence for a sin($\phi/2$)-component in the junction current-phase relation carried by the Majorana bound states that nucleate at the same locations.  The magnitude of the node-lifting is predicted to be independent of a top-gate voltage, suggesting that the critical current at the expected node location comes from the topological states.  We also expect kinks in the diffraction patterns at the magnetic fields at which Josephson vortices enter the junctions, further evidence for these current-carrying Majorana states. As with other systems exhibiting low-lying dispersive states\cite{}, direct measurements of the Josephson current-phase relation would show a pronounced skewness. This feature indicates higher-order harmonics arising from high-transparency states that should dominate near locations in the junction where the phase difference is near an odd-multiple of $\pi$ and the Andreev bound state energy gap closes. We note that many of these features have already been observed and we are preparing a comprehensive review of the current status of experiments \cite{ExperimentPaper}.

It is important to emphasize that none of these observations is conclusive evidence for Majorana modes, which only demonstrating braiding induced parity transitions with non-Abelian statistical behavior can provide. However, analogous to the avidly studied zero-bias conductance peaks in nanowire system, these observations would constitute a first major step. They would pave the way to the braiding schemes we propose that are designed to provide a definitive verification of Majorana  states.

\section{Conclusion}
In this work we have proposed the S-TI-S junctions as a platform for realisation and manipulation of Majorana bound states. While nanowire systems have enjoyed about a decade's worth of attention \cite{Lutchyn18}, there have been several theoretical and experimental efforts in establishing  S-TI-S systems as a viable alternative to realise topological superconductivity\cite{Fu08,Fu09,Potter13, Sochnikov13,Sochnikov15}. We have shown in this work that S-TI-S junctions offer several advantages stemming from extended Josephson junction physics. Manipulation of the superconductor phase difference profile is a natural component in extended Josephson junction systems. Associated magnetic fields are of much lower magnetic fields compared to current requirements for realizing topological superconducting nanowires. Well-established phase-sensitive measurements, such as those using SQUIDs, form versatile probes of the system. The technique of magnetic field piercing through the junctions, the key to nucleating vortices which harbor MBSs, is standard practice evolved over decades for conventional superconductors.  These vortices have added advantage of mobility. By meticulously engineered protocols of applying currents, voltages and local magnetic fields, the motion of MBSs can be realised in controlled ways. 

Here, building on these excellent capabilities of extended topological Josephson junctions, we have propose a series of steps to demonstrate the feasibility of these platforms for performing MBS-based exchange braiding and quantum operations. As parallels to proposals in nanowire systems, we have identified two kinds of junction geometries for performing non-Abelian operations in the Hilbert space of parity qubit states associated with four MBSs. First, the triunctions are a viable alternative to T-junctions\cite{Alicea11} for performing the exchange braiding. Second, effective braiding/fusion of MBSs can be achieved in linear junctions threaded by perpendicular magnetic fields. By tuning the local magnetic field, the coupling between neighbouring Majorana modes can be tuned and this coupling can be used to perform an effective non-Abelian rotation.  We have also mentioned a few accompanying schemes for initializing parity states and performing readouts, essential for quantum processing. They range from observing characteristic Fraunhofer patterns in the Josephson junction to quantum-dot sensing to transmon-based measurements. While the lofty goal of performing topological quantum computational operations may not be accessible in the near future, a proof-or-principle Josephson junction measurement suggesting the existence of MBSs, we find, is within sight.

In conclusion, we have presented an extended topological Josephson junction based platform for nucleation of MBSs and their manipulation towards an eye for quantum processing. Our proposed platform provides an alternative for those based on topological superconducting nanowires; as with conventional qubits, successful realization of quantum computational schemes and quantum information processing would undoubtedly require investigating multiple routes. The platform is amenable to scaling beyond the first steps of nucleating and non-Abelian rotations proposed here. These steps would include measurement of non-local Majorana correlations, integrating multiple circuits and qubits, and forming hybrid systems coupling to conventional qubits, given that MBSs alone cannot form a universal quantum computer. As with nanowire systems, it would be crucial to recognize exactly which circumstances enjoy topological protection and which ones are prone to dissipation and decoherence, be it from quasiparticle poisoning, finite temperature effects, or other factors. While these are all longer-term goals, unequivocal evidence of the existence of Majorana fermions as ascertained by their braiding properties seems to be well within reach in the near future, given the characteristic zero-bias conductance peak signatures in nanowires and signatures of $4\pi-$periodicity of the current-phase relation seen in the critical diffraction patterns in extended Josephson junctions. At the fundamental level, with Majorana's initial postulation of a particle being its own antiparticle dating back to the 1930s, while at the quasiparticle level as opposed to the elementary particle level, a realization of such a quantum entity is a tour de force step in and of itself. 

\begin{acknowledgments}
 We are thankful to insightful discussions with Liang Fu, Alexey Bezryadin, James Eckstein and Can Zhang. This work is supported by the National Science Foundation under grants DMR-1745304 EAGER:BRAIDING (SH, EH, SV), and DMR-1610114 and DMR-2004825 (GY, DVH, SV), and by the Gordon and Betty Moore Foundation’s EPiQS Initiative through Grant GBMF4302 (YW).

\end{acknowledgments}

\bibliographystyle{apsrev}
\bibliography{main}

\end{document}